\documentstyle[psfig,graphicx]{mn}                       
\newcommand{\reduceme}{\mbox{R\raisebox{-0.35ex}{E}D%
\hspace{-0.05em}\raisebox{0.85ex}{uc}\hspace{-0.90em}%
\raisebox{-.35ex}{{m}}\hspace{0.05em}E}}

\title[The near-IR Ca\,{\sc ii} triplet: stellar library]
{Empirical calibration of the near-IR Ca\,{\large \bf II} triplet --
I. The stellar library and index definition}

\author[A.J. Cenarro et al.]  
    {A.J.~Cenarro,$^1$\thanks{E-mail: cen@astrax.fis.ucm.es}
    N.~Cardiel,$^1$ J.~Gorgas,$^1$ R.F.~Peletier,$^2$ A.~Vazdekis,$^3$
\newauthor and F.~Prada.$^4$\\ $^1$Dept. de Astrof\'{\i}sica, Fac. de
    Ciencias F\'{\i}sicas, Universidad Complutense de Madrid, 28040 Madrid,
    Spain.\\ $^2$School of Physics and Astronomy, University of Nottingham,
    University Park, Nottingham NG7 2RD, UK\\ $^3$Dept. of Physics, University
    of Durham, South Road, Durham DH1 3LE, UK\\ $^4$Observatorio de Calar
    Alto, Almer\'{\i}a, Spain\\}
\date{}
\pagerange{\pageref{firstpage}--\pageref{lastpage}}
\pubyear{}

\def\LaTeX{L\kern-.36em\raise.3ex\hbox{a}\kern-.15em
    T\kern-.1667em\lower.7ex\hbox{E}\kern-.125emX}

\begin{document}

\label{firstpage}

\maketitle

\begin{abstract}

A new stellar library at the near-IR spectral region developed for the
empirical calibration of the Ca\,{\sc ii} triplet and stellar
population synthesis modeling is presented. The library covers the
range $\lambda\lambda$~8348-9020~\AA\ at 1.5~\AA\ (FWHM) spectral
resolution, and consists of 706 stars spanning a wide range in
atmospheric parameters.  We have defined a new set of near-IR indices,
CaT$^{*}$, CaT and PaT, which mostly overcome the limitations of
previous definitions, the former being specially suited for the
measurement of the Ca\,{\sc ii} triplet strength corrected for the
contamination from Paschen lines. We also present a comparative study
of the new and the previous calcium indices, as well as the
corresponding transformations between the different systems. A
thorough analysis of the sources of index errors and the procedure to
calculate them is given. Finally, index and error measurements for the
whole stellar library are provided together with the final spectra.
\end{abstract}


\section{Introduction}

With this paper we start a series dedicated to the empirical
calibration of the near-infrared Ca\,{\sc ii} triplet. The ultimate
aim of this work for us is to use the strength of the Ca lines in this
spectral range to investigate the stellar content of early-type
galaxies. However, we expect that researchers in different areas
(e.g. starburst and active galaxies, globular clusters, and stellar
astrophysics) can make use of some of the results and information that
will be presented throughout these papers. These include a new stellar
library, a set of homogeneous atmospheric parameters, empirical
fitting functions describing the behaviour of the Ca triplet with the
stellar parameters, and stellar population model predictions.

Traditionally, elliptical galaxies have been thought to be a uniform
class of objects, with global properties changing smoothly with mass
and hosting old and coeval stellar population. However, over the last
decade, a growing body of evidence is indicating that the formation
processes and star formation histories of, at least, an important
fraction of early-type galaxies is much more complex and
heterogeneous. The apparent age spread among elliptical galaxies
(Gonz\'{a}lez 1993; Faber et al. 1995; J{\o}rgensen 1999), the
distribution of element abundances (Worthey 1998; Peletier 1999;
Trager et al. 2000a) and the interpretation of the scaling relations
(like the colour--magnitude or Mg$_2$--$\sigma$ relations; Bower,
Lucey \& Ellis 1992; Bender, Burstein \& Faber 1993; Pedraz et
al. 1999; Terlevich et al. 1999; Kuntschner 2000; Trager et
al. 2000b), are some of the main issues in the present debate about
the evolutionary status of early-type galaxies. A major common
obstacle to tackle with the previous problems is how to disentangle
age and metallicity effects in the integrated spectrum of a composite
stellar population.

The measurement and interpretation of line-strength indices in the
spectra of early-type galaxies has been a fundamental tool to address
the above topics, the Lick/IDS (Image Dissector Scanner) system
(Worthey et al. 1994 and references therein) being the most widely
used (but see also Rose 1994). The Lick system makes use of strong
spectral features in the blue spectral range (from $\lambda~4100$~\AA\
to $\lambda~6300$~\AA), although only the main absorption lines in the
narrower $\lambda\lambda$~4800--5400~\AA\ range have been extensively
measured with high precision in the spectra of early-type
galaxies. More recently, new spectral indices have been defined and
calibrated in the bluer region (Jones \& Worthey 1995; Worthey \&
Ottaviani 1997; Vazdekis \& Arimoto 1999; Gorgas et al. 1999), with
enlarged capability to decouple age and metallicity effects. However,
the spectral range of interest is still quite narrow. It is obvious
that enlarging the wavelength coverage will allow us to investigate
the relative contributions of different stellar types to the composite
spectrum. As an example, a comparison between the mean ages derived
using, as an age discriminant, features in different spectral regions,
can provide important constraints about the star formation history of
the galaxies (S\'{a}nchez--Bl\'{a}zquez et al. 2000). Although an
important effort has been made to extend the stellar population
studies to the ultraviolet region (e.g. Ponder et al. 1998), this is
not the case for the near-infrared. In fact, it is remarkable that,
although the features in the near-infrared spectral range where
already included in the early analyses of the extragalactic old
stellar populations (e.g. Spinrad \& Taylor 1971), 30 years later, and
despite the advent of modern CCD detectors, the potential of this
spectral range to investigate the stellar population of early-type
galaxies is still almost unexploited.

The Ca\,{\sc ii} triplet is one of the most prominent features in the
near-IR spectrum of cool stars and its potential to study the
properties of stellar populations has been extensively acknowledged in
the literature (see Section~\ref{prevref2} for a review of previous
works on the subject). However, a reliable analysis of the Ca triplet
measurements in integrated spectra rests on the comparison of the data
with the predictions of stellar population models. The accuracy of
such predictions is highly dependent on the input calibration of the
calcium line-strengths in terms of the main atmospheric stellar
parameters (namely effective temperature, surface gravity and
metallicity). The quality of this calibration has been the major
drawback of previous stellar population models which have included
predictions for the strength of the calcium triplet (Vazdekis et
al. 1996; Idiart, Th\'{e}venin \& de Freitas Pacheco 1997, hereafter
ITD; Mayya 1997; Garc\'{\i}a--Vargas, Moll\'{a} \& Bressan 1998;
Schiavon, Barbuy \& Bruzual 2000; Moll\'{a} \& Garc\'{\i}a--Vargas
2000).  Previous calibrations have been either theoretical (based on
model atmospheres), with their associated uncertainties, or based on
empirical stellar libraries with a poor coverage of the atmospheric
parameter space.  Apart from this major problem, there are other
factors which have compromised, with different impacts, the
reliability of these previous papers. They include stellar libraries
with too few stars, problematic index definitions (e.g., they cannot
be used for all spectral types), uncertainties in the input stellar
parameters (which translate into unknown uncertainties in the derived
predictions), and illnesses in the fitting procedures (a proper
statistical analysis of the computed coefficients is seldom
followed). Throughout this work we will review and comment in detail
all these problems and how we have coped with them.

The main motivation of the present series of papers is, therefore, to
provide a reliable calibration of the near-infrared Ca\,{\sc ii}
triplet that makes it possible to accurately calculate the behaviour
of the calcium line-strengths in stellar populations with a wide range
of ages and metallicities. The outline of the series is as follows. In
this first paper we introduce the new stellar library and define a new
set of indices to quantify the strength of the Ca triplet. Paper~II is
dedicated to the determination of the input atmospheric parameters for
the stellar sample. These are some of the basic ingredients of the
empirical fitting functions and the spectral synthesis analysis that
will be presented in Paper~III and~IV, respectively. In this latter
paper, we will compute the predictions for single-burst stellar
population models in two parallel ways: providing the Ca triplet
strength using the fitting functions of Paper~III, and supplying
spectral energy distributions in the $\lambda\lambda$~8348--9020~\AA\
range (following Vazdekis 1999).

In Section~\ref{prevref} of this paper we review previous work on the
Ca\,{\sc ii} triplet behaviour and its applications to different
astrophysical problems.  Section~\ref{stellib} describes the new
stellar library as well as an overview of the observation and
reduction procedures. A thorough discussion of the Ca\,{\sc ii} index
definition is given in Section~\ref{indexdef}, where we define new
improved Ca\,{\sc ii} indices which overcome the problems affecting
previous definitions.  This section also includes a comparative study
of the sensitivity of the Ca\,{\sc ii} indices to effects like the
signal-to-noise ratio or the spectral resolution, and a set of
calibrations for converting between different index definitions.  In
Section~\ref{errors}, an extensive analysis of index errors is
presented, discussing the different sources of random errors and
systematic effects. As a supplement of this section, Appendix~A
provides accurate formulae for the computation of random errors in the
index measurements, together with an analytical estimate of errors
from signal-to-noise ratios. Finally, in Section~\ref{database} we
refer the reader to a web page in which we provide a database
containing the spectra for the whole library, an electronic table
listing full information for each star, and public {\sc fortran}
routines to compute the new indices and their associated errors.

\section{Previous works on the near-IR Ca\,{\sc ii} triplet} 
\label{prevref}

In a pioneer work, Merrill (1934) (see also Wilson \& Merrill 1937)
explored for the first time the near-infrared region of stellar
spectra, identifying the most relevant absorption features and noting
that {\sl the lines of Ca\,{\sc ii} $\lambda\lambda$~8498, 8542, 8662
are prominent in types A--M and may have an interesting relationship
to absolute magnitude\/}. This tentative prediction was confirmed by
the subsequent works of Keenan \& Hynek (1945) and Parsons (1964). In
particular, this last author showed how the strength of the Ca lines
(blended with Paschen lines for early--type stars) increases with
absolute magnitude (i.e. being larger for supergiants) at fixed
spectral type. Later works by O'Connell (1973), Anderson (1974), Cohen
(1978), and others (see below) confirmed these findings, establishing
the near-infrared Ca triplet as one of the most striking
luminosity-sensitive feature in the spectra of F--M stars.

\subsection{Calibrations of the calcium line-strengths}
\label{prevref1}

After these early studies, several authors have tried to model the
behaviour of the Ca triplet with the basic atmospheric parameters
($T_{\rm eff}$, $\log g$ and [Fe/H]). Two main lines of approach have
been followed, either using empirical stellar libraries, or using the
predictions of stellar atmosphere models.

\subsubsection{Empirical calibrations}
\label{empical}

Among the works following the empirical method, we must remark on that
of Jones, Alloin \& Jones (1984, hereafter JAJ), in which they found
that the Ca triplet strength strongly correlates with gravity (in the
sense that it increases for giant and supergiant stars). They also
noted that the residuals of this relation correlate weakly with
metallicity. This strong gravity dependence was independently
confirmed by Carter, Visvanathan \& Pickles (1986) and Alloin \& Bica
(1989), although they did not agree in their conclusions about the
metallicity dependence, being stronger in the latter study.

An important step to quantify the $g$ and [Fe/H] dependence was given
by the comprehensive study of D\'{\i}az, Terlevich \& Terlevich (1989,
hereafter DTT). Using a stellar library with an enlarged range in
gravity and metallicity, these authors quantified the biparametric
behaviour with $\log g$ and [Fe/H], noting that, in the
high-metallicity range, the strength of the Ca triplet depends only on
gravity, whilst, for low-metallicity stars, [Fe/H] is the main
parameter.  Zhou (1991, hereafter ZHO) carried out a similar analysis,
using a higher spectral resolution (2~\AA) and extending the sample of
stars by including cooler stars (up to M7). He found that the effect
of temperature, found to be negligible in the previous work, was
important at low $T_{\rm eff}$. He also noted that the gravity
dependence was stronger for giants than for dwarfs.

Another relevant study is that of Mallik (1994, 1997). Through the
analysis of an ample stellar library at high resolution (0.4~\AA),
this author confirmed the strong dependence on $\log g$ (and showed
that it was stronger as metallicity increases) and the milder
dependence on [Fe/H] (but, remarkably, more important for supergiant
stars, in agreement with DTT). Unfortunately, he does not provide any
fitting function, restricting the potential application of these
qualitative results. ITD have derived empirical functions which
predict the strength of the Ca triplet in terms of the three
atmospheric parameters. Using a library with a good coverage in
metallicity but lacking a representative sample of supergiants, they
conclude that the gravity dependence is not as strong as previously
reported, and that metallicity is the main parameter.

It is clear that there are some apparent contradictions among the
conclusions of these papers. A critical analysis of these will be
presented in Paper~III of this series, where we will show that the
apparent discrepancies are mainly due to the different ranges of
stellar parameters in the employed stellar libraries, the diversity of
index definitions (see Section~\ref{previndex}), and differences in
the fitting procedures (e.g. in most cases, a proper statistical
approach was not applied).

\subsubsection{Theoretical calibrations}

The theoretical modelling of the Ca triplet lines is not simple since
the line cores are formed in the stellar chromosphere and, therefore,
non--Local Thermodynamic Equilibrium (NLTE) models are required. For
this reason, the first attempts to analyze the sensitivity of the Ca
strength to the atmospheric parameters were restricted to the line
wings. In this sense, Smith \& Drake (1987, 1990) concluded that the
dependence on metallicity should be larger than that reported by the
empirical studies (like JAJ), and that effective temperatures had a
non negligible effect. Erdelyi--Mendes \& Barbuy (1991), also using
LTE models, extended the analysis to cooler stars and included the
significant contribution of molecular bands. They agreed with Smith \&
Drake in that metallicity, rather than gravity, was the main
parameter, the [Fe/H] dependence being stronger for lower gravities.

Using NLTE models, J{\o}rgensen, Carlsson \& Johnson (1992) carried
out, for the first time, the computation of the line-cores, deriving
full equivalent-widths (they showed, however, that the effects of
departures from LTE are negligible since the equivalent widths are
dominated by the line wings). Concerning the sensitivity to stellar
parameters, they found that the gravity dependence of the Ca triplet
is influenced by $T_{\rm eff}$ and [Fe/H] in a complicated way, They
provided fitting functions for stars in the range 4000~K $\leq T_{\rm
eff}\leq$ 6600~K and $-1.0\leq{\rm [A/H]}\leq+0.2$~dex.

Recently, Chmielewski (2000) has presented a comprehensive analysis of
the theoretical modelling of the Ca lines. He stresses that previous
works did not take into account the contribution of hydrogen Paschen
lines to the integrated equivalent widths, showing that this
contribution is significant for dwarfs hotter than 5800~K and giants
with $T_{\rm eff}$ above $\sim 5500$~K. Interestingly, this effect
explains most of the discrepancies between the previous empirical and
theoretical works concerning the temperature dependence of the Ca
lines. His main conclusions about the Ca line-strength sensitivity to
the stellar parameters are that the gravity dependence only holds for
$\log g<3$~dex, metallicity is a significant parameter in all cases,
and temperature effects can only be neglected for dwarfs between 5000
and 6000~K.

Unfortunately, these theoretical studies, being very important to help
to understand the behaviour of the Ca triplet and to check the
empirical studies, are restricted to cool stars (typically between F
and K) and, obviously, their conclusions can not be extrapolated to
later spectral types to construct stellar population synthesis models
of, relatively old, stellar systems.

\subsection{Applications of the Ca triplet to the study of stellar systems}
\label{prevref2}

Over the years, the near--IR Ca triplet has been used extensively to
address a number of topics in different areas.

\subsubsection{Stellar astrophysics}

In the field of stellar astrophysics, the triplet, together with other
spectral features in this spectral range, has been used for the
spectral classification of stars (Sharpless 1956; Bouw 1981;
Kirkpatrick, Henry \& McCarthy 1991; Ginestet et al. 1994; Munari \&
Tomasella 1999). These features have also been widely used as
chromospheric activity indicators (e.g. Linsky et al. 1979; Dempsey et
al. 1993; Montes et al. 1998). Furthermore, its sensitivity to surface
gravity has been exploited to identify supergiants in the Galaxy
(e.g. Garz\'{o}n et al. 1997) or in Local Group galaxies (Mantegazza
1992; Humphreys et al. 1988; Massey 1998).

\subsubsection{Globular clusters}

In spite of the first empirical papers that remarked that the Ca
triplet was mainly gravity driven (see above), Armandroff \& Zinn
(1988, hereafter A\&Z), in a pioneering study, found that, in the
integrated light of galactic globular clusters, the Ca equivalent
width was well correlated with metallicity, indicating that this
feature could be a fair metallicity indicator for old, and
approximately coeval, stellar populations, provided that the
metallicity was lower than solar.

Subsequent works in this field concentrated in the study of individual
red giants of globular clusters. Armandroff \& Da Costa (1991,
hereafter A\&D) derived an empirical relation between the cluster
metallicity and the ``reduced'' equivalent width $W^\prime=EW({\rm
CaT})+c\ (V-V_{\rm HB})$, where $c$ is a constant term, $EW({\rm
CaT})$ is the pseudo--equivalent width (typically the sum of the two
strongest lines of the triplet), and $V_{\rm HB}$ is the $V$ magnitude
of the horizontal branch. Note that $W^\prime$ is a monotonic function
of metallicity because gravity effects are removed by the last term of
the equation. This relation has been recalibrated by several authors
(e.g. Da Costa \& Armandroff 1995; Geisler et al. 1995; Rutledge et
al. 1997ab, hereafter RHS) and extensively used to derive
metallicities of galactic globular clusters (Armandroff, Da Costa \&
Zinn 1992; Da Costa, Armandroff \& Norris 1992; Buonanno et al. 1995;
Suntzeff \& Kraft 1996; Rosenberg et al. 1998), clusters and
individual stars in the Magellanic Clouds (Olszewski et al. 1991;
Suntzeff et al. 1992; Da Costa \& Hatzidimitriou 1998; Cole,
Smecker--Hane \& Gallagher 2000), and Local Group dwarf spheroidals
(Suntzeff et al. 1993; Smecker--Hane et al. 1999). A comprehensive
analysis of the different methods to measure cluster metallicities
using the the Ca triplet can be found in RHS.

\subsubsection{Active galaxies and extragalactic H\,{\sc ii} regions}

One of the fields in which the study of the Ca triplet has had a wide
application is that of active galaxies and starburst
regions. Terlevich, D\'{\i}az \& Terlevich (1990a) studied the Ca
triplet in a sample of normal and active galaxies, concluding that
LINERS and Seyfert 2 objects exhibit Ca strengths equal or larger than
those found in normal ellipticals (in contrast with the blue
absorption lines, which are usually diluted in active
galaxies). Following the conclusion of DTT that Ca equivalent widths
above 9~\AA\ are only found in red supergiant stars, they interpret
this result as a strong evidence for the occurrence of nuclear
starbursts in the central regions of active galaxies. This same
approach has been followed by a number of authors (Forbes, Boisson \&
Ward 1992; Garc\'{\i}a--Vargas et al. 1993; Gonz\'{a}lez Delgado \&
P\'{e}rez 1996ab; Heckman et al. 1997; P\'{e}rez et al. 2000) to study
the evolutionary status of the nuclei and circumnuclear regions of
active galaxies through the presence of red supergiants.

The Ca triplet has also been detected in extragalactic H\,{\sc ii}
regions (Terlevich et al. 1990b, 1996), being interpreted as due to
the recent star formation burst, and it has been used to study the
star formation history of extragalactic super-star clusters (Prada,
Greve \& McKeith 1994; Garc\'{\i}a--Vargas et al. 1997; Gonz\'{a}lez
Delgado et al. 1997; Gallagher \& Smith 1999; Goudfrooij et al. 2001).

\subsubsection{Early-type galaxies}

The first papers analyzing the Ca triplet in early-type galaxies tried
to make use of its gravity sensitivity to constrain the dwarf/giant
ratio in the integrated light of old stellar populations. The main
debate at that time was focussed on the possible dwarf enrichment in
the nucleus of M~31, as compared to its bulge or to low luminosity
ellipticals like M~32.  While Faber \& French (1980) and, to some
degree, Carter et al. (1986) favored a dwarf-enriched stellar
population, Cohen (1978) and Alloin \& Bica (1989) found that
metallicity effects and contamination by a molecular band could
account for the apparent enrichment, mainly based on measurements of
the strength of the Na\,{\sc I} feature at $\lambda$~8190~\AA.

One interesting result from these first studies was that it was found
that the strength of the Ca triplet did not vary much among early-type
galaxies (Cohen 1979). Note that although Cohen (1978) found that the
nucleus of M~31 exhibited stronger Ca lines than M~32, Faber \& French
(1980) reported the opposite behaviour. Later, Bica \& Alloin (1987,
hereafter B\&A) measured the Ca strength in the spectra of 62 galactic
nuclei (from E to Sc), concluding that the equivalent widths were not
related to galaxy types or luminosities. This was confirmed by the
studies of Terlevich et al (1990a), in which they found a small spread
in the Ca strengths of 14 normal galaxy nuclei, and Houdashelt (1995),
who noted that the Ca equivalent widths of a sample of 34 early-type
galaxies did not vary significantly among galaxies of different color
or absolute magnitude.

The apparent uniformity of the Ca measurement in elliptical galaxies
is somewhat surprising since: i) It contrasts with the metallicity
dependence found for the stellar samples and the reported behaviour in
galactic globular clusters (see above), ii) Previous stellar
population models predict a high sensitivity of the Ca triplet to the
metallicity of metal-rich old stellar populations (Garc\'{\i}a--Vargas
et al. 1998; Schiavon et al. 2000), and iii) It has been argued that,
in contrast to other $\alpha$-elements, Ca abundance is not enhanced
compared to Fe in bright elliptical galaxies (O'Connell 1976; Vazdekis
et al. 1997; Worthey 1998; Moll\'{a} \& Garc\'{\i}a--Vargas 2000), and
variations of the Fe line-strengths among ellipticals, although
difficult to detect, are not negligible (e.g. Gorgas, Efstathiou \&
Arag\'{o}n--Salamanca 1990; Gonz\'{a}lez 1993; Davies, Sadler \&
Peletier 1993; Kuntschner 2000). Therefore, more reliable predictions
of the Ca triplet behaviour in the integrated stellar population of
high metallicity systems and new high-quality measurements are hardly
needed to clarify this point.

This problem is closely related to the radial behaviour of the Ca
strengths within galaxies. Given the measured metallicity gradients
usually found in elliptical galaxies (see Gonz\'{a}lez \& Gorgas 1996,
and references therein), even using Fe line-strengths, one should
expect to find unambiguous negative gradients in Ca. However, Cohen
(1979) and Boroson \& Thompson (1991) found negligible gradients;
Carter et al (1986) and Peletier et al. (1999) measured either null or
positive gradients, and Delisle \& Hardy (1992, hereafter D\&H) found
negative gradients in many galaxies of their sample.

Another important difficulty arises when trying to compare the
absolute values of the Ca triplet measured in early-type galaxies with
the actual values predicted by stellar population synthesis models
(e.g. Peletier et al. 1999; Moll\'{a} \& Garc\'{\i}a--Vargas 2000).
The uncertainties of the theoretical or empirical Ca behaviour
implemented in such models, together with other problems such as
observational uncertainties, corrections from velocity dispersion or
transformations between different index definitions, make very
difficult to extract any significant conclusion from such a
comparison.

\section{The near-IR stellar library}
\label{stellib}

\subsection{Previous libraries in the near-IR spectral range}

Previous stellar libraries providing spectra in the Ca\,{\sc ii}
triplet spectral range are listed in Table~\ref{tab_lib}. Here we only
include libraries with a spectral resolution better than 10~\AA\
(FWHM). For a list of lower resolution libraries we refer the reader
to Munari \& Tomasella (1999).  We have also excluded from this list
some libraries which are too specific for the purposes of this work,
like those of Fluks et al. (1994) (only M stars), Allen \& Strom
(1995) (stars from two open clusters), RHS (huge collection of red
giants from galactic globular clusters), or Montes \& Mart\'{\i}n
(1998) (high resolution library in a limited spectral range). It is
clear from this table that no previous library provides simultaneously
a broad coverage of stellar metallicities and effective temperatures.
In fact, the only libraries including a good fraction of low
metallicity stars are those of DTT and ITD, but the spectral type
ranges spanned in these cases are rather limited, especially in the
latter case (as we will see in Paper~IV, the relative contribution of
stars colder than K3 to the near-IR spectrum of an old stellar
population is very important). On the other hand, libraries
composed of stars of all spectral types do not attain a broad range in
metallicity (e.g. Munari \& Tomasella 1999). The low frequency of
supergiant stars is an additional problem of some libraries (JAJ;
ITD). It must be noted that some of the apparent discrepancies among
different authors who have modelled the behaviour of the CaT index
with the atmospheric parameters are mainly due to this different
coverage of the parameter space by the calibrating stars (see
Paper~III).

\begin{table*}
\begin{minipage}{160mm}
\caption{Medium-resolution stellar libraries in the Ca\,{\sc ii} triplet 
spectral region.}
\label{tab_lib}
\begin{tabular}{lllrrl}
\hline
Reference  &  \multicolumn{1}{c}{No. stars} & \multicolumn{1}{c}{Resolution} &
\multicolumn{1}{c}{Spectral}  &  \multicolumn{1}{c}{[Fe/H]}    &        
Comments\\
           &      & \multicolumn{1}{c}{(FWHM, \AA)}&  
\multicolumn{1}{c}{types}     & \multicolumn{1}{c}{range}   &      \\\hline
Jones et al. (1984)        & \makebox[23pt][r]{62}  & \makebox[22pt][r]{3}   & 
B -- M5 & $-0.60$ , $+0.23$&\\ 
D\'{\i}az et al. (1989)    & \makebox[23pt][r]{106} & \makebox[22pt][r]{3}.5 & 
F5 -- M1 & $-2.70$ , $+0.55$&\\
Zhou (1991)                & \makebox[23pt][r]{144} & \makebox[22pt][r]{2}   & 
F5 -- M7 & $-2.28$ , $+0.60$& 2 stars with [Fe/H]$<-0.61$\\
Andrillat et al. (1995)    & \makebox[23pt][r]{76}  & \makebox[22pt][r]{1}.2 & 
O5 -- G0 & \multicolumn{1}{c}{?}&\\
Serote Roos et al. (1996)  & \makebox[23pt][r]{21}  & \makebox[22pt][r]{1}.25& 
B3 -- M5 & $-0.15$ , $+0.39$& Only giants and supergiants\\
Carquillat et al. (1997)   & \makebox[23pt][r]{36}  & \makebox[22pt][r]{2}   & 
F5 -- M4 & \multicolumn{1}{c}{?}&\\
Idiart et al. (1997)       & \makebox[23pt][r]{55}  &\makebox[22pt][r]{$\sim2$}&     
A1 -- K3 & $-3.15$ , $+0.35$&\\
Mallik (1997)              & \makebox[23pt][r]{146} & \makebox[22pt][r]{0}.4 &
F7 -- M4 & $-3.0$ , $+1.01$& 1 stars with [Fe/H]$<-1.6$\\
Munari \& Tomasella (1999) & \makebox[23pt][r]{131} & \makebox[22pt][r]{0}.43& 
O4 -- M8 & $-0.54$ , $+0.30$&\\
Montes et al. (1999)       & \makebox[23pt][r]{130} & \makebox[22pt][r]{$\sim0$}.7 &     
F0 -- M8 & $-2.74$ , $+0.31$& 3 stars with [Fe/H]$<-0.5$\\
This work                  & \makebox[23pt][r]{706} & \makebox[22pt][r]{1}.5 & 
 O6 -- M8 & $-3.45$ , $+0.60$&\\
\hline
\end{tabular}
\end{minipage}
\end{table*}

\subsection{The sample}

We have observed a new stellar library of 706 stars at the
near-infrared spectral region ($\lambda\lambda$~8348-9020~\AA). It
includes 421 of the 424 stars with known atmospheric parameters of the
Lick/IDS Library (Burstein et al. 1984; Faber et al. 1985; Burstein,
Faber \& Gonz\'{a}lez 1986; Gorgas et al. 1993; Worthey et
al. 1994). This subsample spans a wide range in spectral types and
luminosity classes. Most of them are field stars from the solar
neighbourhood, but stars covering a wide range in age (from open
clusters) and with different metallicities (from galactic globular
clusters) are also included. In order to obtain a large sample of
stars in common with other previous works devoted to the Ca\,{\sc ii}
triplet, our stellar library was enlarged to include 105 of the 106
stars of DTT and 43 of the 55 stars of the sample of ITD. Moreover,
with the aim of filling gaps in the parameter space, we also included
stars from several other compilations, mainly focusing on O, B and A
types (37 hot stars from the sample of Andrillat, Jaschek \& Jaschek
1995), late M types, metal poor, metal rich and chemically peculiar
stars (40 stars from a list kindly provided by G. Worthey, private
communication, and 102 stars from Jones 1997). The presence of hot
stars in the stellar library allows us to expand the predictions of
our models to young stellar populations. Also, and as it will be shown
in Paper~IV, very cool stars are necessary to reproduce the integrated
spectra of old stellar systems at the near-infrared spectral region
over a wide range of ages. Finally, since calcium is an
$\alpha$-element, 29 additional stars with high or low Ca/Fe ratios
from the catalogue by Th\'evenin (1998) were observed in order to
analyze the Ca\,{\sc ii} index dependence on relative abundances (see
Paper~III).

The new stellar library covers the following ranges in atmospheric
parameters: {\it T}$_{{\rm eff}}$ from 2750 to 38400~K, $\log g$ from
0 to 5.12~dex, and [Fe/H] from $-3.45$ to $+0.60$~dex. It is important
to note that the stars of the stellar library do not homogeneously
cover the parameter space. In fact, most of the stars ($\sim$ 83 per
cent) have metallicities from $-1.0$ to $+0.5$~dex, and the widest
gravity range is also within this metallicity interval. The final
atmospheric parameters which have been adopted for each star and the
method to derive them are discussed in detail in Paper~II of this
series. We refer the reader to that paper for an HR diagram of the
whole sample.

\subsection{Observations and data reduction}
\label{obsreduc}

\begin{table*}                                                              
\centering{                                                                
\caption{Observational configurations}          
\label{instrum}                                                            
\begin{tabular}{crcccccccr}          
\hline                                
Run & Dates\ \ \ \ \  & Telescope & Spectrograph  & Detector  & Dispersion & 
$\Delta\lambda$ & Slit width & FWHM  & $N_{\rm obs}$\\
    &                 &           &               &           & (\AA/pix)  & 
(\AA)           & (arcsec) & (\AA) &     \\
\hline
1 & 20--26 Sep 1996 & JKT 1.0 m  & RBS          & TEK$\#$4   & 0.85 & 8331--9200 & 1.5 & 1.18 & 264 \\
2 & 15--19 Jan 1997 & JKT 1.0 m  & RBS          & TEK$\#$4   & 0.85 & 8331--9200 & 1.5 & 1.50 & 248 \\
3 & 25--29 Jun 1997 & JKT 1.0 m  & RBS          & TEK$\#$4   & 0.85 & 8331--9200 & 1.5 & 1.07 & 362 \\
4 & 13 Aug 1997     & INT 2.5 m  & IDS          & TEK$\#$3   & 0.85 & 8185--9055 & 1.5 & 1.28 &  33 \\
5 & 4 Aug 1996      & WHT 4.2 m  & ISIS         & TEK$\#$2   & 0.79 & 8222--9031 & 1.2 & 2.13 &  12 \\
6 & 17--18 Nov 1996 & CAHA 3.5 m & TWIN         & SITe$\#$4d & 0.81 & 8300--9912 & 2.1 & 2.11 &  15 \\
\hline
\end{tabular}                                                              
}                                                                          
\end{table*}

The spectra of the stellar library were obtained during a total 21
nights in six observing runs from 1996 to 1997 using the JKT (Jacobus
Kapteyn Telescope), INT (Isaac Newton Telescope) and WHT (William
Herschel Telescope) at the Roque de los Muchachos Observatory (La
Palma, Spain) and the 3.5-m telescope at Calar Alto Observatory
(Almer\'{\i}a, Spain). It is important to note that most of the stars
($\sim$ 93 per cent), including all the field stars and the brightest
ones from open and globular clusters, were observed during the three
runs at the JKT using the same instrumental configuration. This
ensures a high homogeneity among the data of these three runs (see the
analysis in Section~\ref{sec33}). The instrumental configuration at
the other telescopes was also chosen to be as similar as possible to
that at the JKT, obtaining spectral resolutions (FWHM) for the six
runs in the range from 1.07~\AA\ to 2.13~\AA. A full description of
these and other instrumental details is given in
Table~\ref{instrum}. We must note that, as it will be explained in
Section~\ref{systematic}, spectra from runs 1, 3 and~4 were broadened
to the spectral resolution of run~2 (1.5~\AA). Hence, and with the
exception of the few stars observed in runs 5 and 6, the whole library
has a common resolution.  $N_{\rm obs}$ stands for the number of stars
observed in each run.

Typical exposure times varied from a few seconds for bright stars to
1800 s for the faintest cluster stars. These ensured typical values of
{\it SN}(\AA) $\sim$ 100~\AA$^{-1}$ (signal-to-noise ratio per
angstrom) for field and open cluster stars, and {\it SN}(\AA) $\geq
15$ for the globular cluster stars.  In order to test the derived
random errors of the final index measurements (see
Section~\ref{random}), we performed multiple observations of a
subsample of stars (15 -- 20) within each particular run and in common
with other different runs.  Finally, to perform a reliable flux
calibration, several (3 or 4) spectrophotometric standards were
observed along each night at different zenithal distances.

The reduction of the data was performed with
\reduceme\,\footnote{http://www.ucm.es/info/Astrof/reduceme/reduceme.html}
(Cardiel 1999), which allows a parallel treatment of data and error
frames (see more details in Section~\ref{errors}) and, therefore,
produces an associated error spectrum for each individual data
spectrum. We carried out a standard reduction procedure for
spectroscopic data: bias and dark subtraction, cosmic ray cleaning,
flat-fielding, C-distortion correction, wavelength calibration,
S-distortion correction, sky subtraction, atmospheric extinction
correction, spectrum extraction, and relative flux calibration. We did
not attempt to obtain absolute fluxes since both the evolutionary
synthesis codes and the line-strength indices only require relative
fluxes. Cluster stars were also corrected from interstellar reddening
using the color excesses from Gorgas et al. (1993) and Worthey et
al. (1994) and the averaged extinction curve of Savage \& Mathis
(1979).

In order to optimize the observing time during the JKT runs, we
decided not to acquire flat-field and comparison arc frames for each
individual exposure of a library star. Concerning the flat-field
correction, after checking that small variations of the CCD
temperature do not affect the flat-field structure and considering
that it exclusively depends on the position of the telescope, we
obtained a complete set of flat-field exposures by pointing at a grid
of positions on the dome (with a resolution of 30\degr\ in azimuth and
15\degr\ in altitude) during the day. Later, using the alt-azimuth
coordinates of each star, the closest flat-field frame over the whole
set was finally used as its own flat-field calibration frame. It is
important to highlight the importance of an accurate flat-field
correction when the fringe pattern becomes relevant, as it is usually
the case in this spectral range. Depending on the properties of the
CCD, the fringe effect introduced high frequency structures with an
amplitude of up to 7 per cent of the true flux, but they virtually
disappeared by using our procedure.  Flux differences between the
normalized fringe patterns of different flat-field frames allows an
estimation of the uncertainties introduced in the fringe correction
procedure. These are always below 0.6 per cent.

Concerning the wavelength calibration, we only obtained comparison arc
frames for a previously selected subsample of stars covering all the
spectral types and luminosity classes in each run. The selected
spectra were wavelength calibrated with their own arc exposures taking
into account their radial velocities, whereas the calibration of any
other star was performed by a comparison with the most similar,
already calibrated, reference spectrum. This working procedure is
based on the expected constancy of the functional form of the
wavelength calibration polynomial within a considered observing run
(which has been widely tested by comparing the derived polynomials for
each run). In this sense, the algorithm that we used is as follows:
after applying a test x-shift (in pixels) to any previous wavelength
calibration polynomial, we obtained a new polynomial which was used to
calibrate the spectrum. Next, the calibrated spectrum was corrected
from its own radial velocity and, finally, the spectrum was
cross-correlated with a reference spectrun of similar spectral type
and luminosity class, in order to derive the wavelength offset between
both spectra. By repeating this procedure, it is possible to obtain
the dependence of the wavelength offset as a function of the test
x-shifts and, as a consequence, we derive the required x-shift
corresponding to a null wavelength offset. Uncertainties in the
wavelength calibration are estimated in Section~\ref{errors}.

It is also important to account for the presence of telluric
absorptions, mainly at the red end of the spectral range (the
strongest lines are locate at $\lambda\lambda$~8952, 8972, 8980,
8992~\AA). Fortunately, since these strong H$_{2}$O lines do not
affect the Ca\,{\sc ii} triplet region and the observations were
performed under dry conditions, we did not correct for their
contamination and it is possible that our final spectra present
unremoved features at the red end. An illustrative example of the
effect of telluric lines in the Ca\,{\sc ii} triplet region is given
in Stevenson (1994) and Chmielewski (2000).

\section{Ca {\sc ii} triplet indices definition}
\label{indexdef}

Line-strength indices have been widely defined to obtain objective
measurements of any relevant spectral feature. For the Ca triplet,
there exist previous index definitions from other authors which were
optimized to measure the lines for a narrow range of spectral types
(mainly G and K-types). These previous indices, however, are not
appropriate for all the spectral types. Hot and cold stars show strong
spectral features which were not taken into account for those index
definitions and, as a consequence, most of the calcium measurements
for these spectral types become unrealistic.  Moreover, some of the
previous definitions have not been optimized for composite stellar
systems. Some of them require spectra with high signal-to-noise ratio,
becoming unpractical for faint objects. Besides, some indices show a
strong dependence on spectral resolution, or velocity dispersion
broadening. In this sense, one should not forget that, in the study of
stellar populations, to measure true equivalent widths is not as
important as to obtain robust index measurements, specially in spectra
with low signal-to-noise ratios or with a wide range of spectral
resolutions.

In order to cope with the above problems, we have decided to introduce
some new Ca\,{\sc ii} triplet index
definitions. Section~\ref{specrange} presents an overview of the
strongest spectral features in the range from 8350 to 9020~\AA. In
Section~\ref{previndex} we analyze the behaviour and limitations of
previous Ca\,{\sc ii} triplet index definitions for different spectral
types. The new indices are defined in Section~\ref{newdefs}, whereas
Section~\ref{sensitivity} is reserved to study the dependence of the
new and old index definitions en the S/N ratio, the combined effect of
spectral resolution and velocity dispersion broadening, and the flux
calibration. Finally, in Section~\ref{calibrations} we present
comparisons and calibrations between different Ca\,{\sc ii} index
systems.

\subsection{The $\lambda\lambda$~8350-9020~\AA\ spectral region}
\label{specrange}

\begin{figure*}
\centerline{\hbox{
\psfig{figure=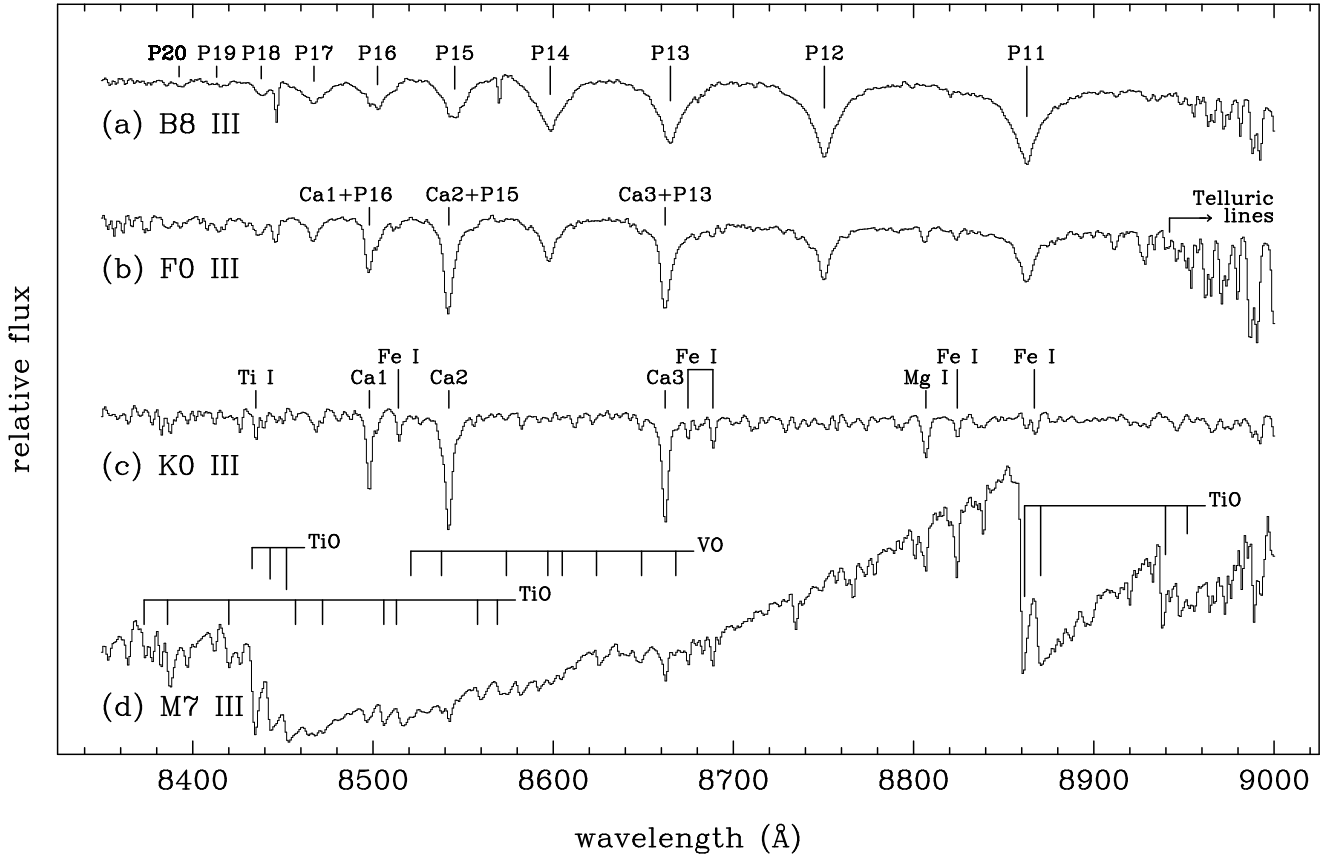}
}} 
\caption{Spectra of the stars HD 186568 (B8 III), HD 89025 (F0 III),
HD 216228 (K0 III) and HD 114961 (M7 III) in the spectral range of the
stellar library. The strongest features in this region are marked: the
Paschen Series (from P11 to P20), the Ca\,{\sc ii} triplet (Ca1, Ca2
and Ca3), several metallic lines (Fe\,{\sc i}, Mg\,{\sc i} and Ti\,{\sc
i}), molecular bands (TiO and VO) and telluric absorptions.}
\label{secLines}
\end{figure*}

Prior to any index definition, we have explored how the main
absorption features in our wavelength range change along the spectral
type sequence (Fig.~\ref{secLines}). Apart from several atomic lines
of intermediate strength, as those of Fe~{\sc i}
($\lambda\lambda$~8514.1, 8674.8, 8688.6, 8824.2~\AA), Mg~{\sc i}
(8806.8~\AA) and Ti~{\sc i} (8435.0~\AA), the Ca\,{\sc ii} triplet
($\lambda\lambda$~8498.02, 8542.09, 8662.14~\AA) is the strongest
feature over a wide range of spectral types (from F5 to M2
approximately) and for all luminosity classes (see
Fig.~\ref{secLines}c).

The hydrogen Paschen series ($\lambda\lambda$~8359.0, 8374.4, 8392.4,
8413.3, 8438.0, 8467.3, 8502.5, 8545.4, 8598.4, 8665.0, 8750.5,
8862.8, 9014.9~\AA, from P22 to P10 respectively) is apparent in stars
hotter than G3 types. Depending on the luminosity class, the strength
of this series reaches a maximum for F or A type stars (Andrillat et
al. 1995). In the earliest spectral types, where the Ca\,{\sc ii}
strength becomes insignificant, the relative depths of the Paschen
lines show a smooth sequence with wavelength (see
Fig.~\ref{secLines}a). However, due to the fact that, for low and
intermediate spectral resolution, the Paschen lines P13, P15 and P16
overlap with the Ca\,{\sc ii} triplet, these three Paschen lines stand
out in the smooth sequence for A and F types
(Fig.~\ref{secLines}b). The fact that the Ca\,{\sc ii} lines are
blended with the Paschen lines has always been an obstacle to measure
the calcium triplet in warm stars (see B\&A and Chmielewsky 2000).

Stars cooler than early M types exhibit molecular bands that change
the slope of the local continuum (Fig.~\ref{secLines}d). The strongest
ones are the triple-headed band at $\lambda\lambda$~8432, 8442,
8452~\AA\ and the double-headed bands at $\lambda\lambda$~8859.6,
8868.5~\AA\ and $\lambda\lambda$~8937.4, 8949.8~\AA\ of titanium oxide
(TiO). Other molecular features are the band sequence of TiO at
$\lambda\lambda$~8373, 8386, 8420, 8457, 8472, 8506, 8513, 8558,
8569~\AA\ and several vanadium oxide (VO) bands at
$\lambda\lambda$~8521, 8538, 8574, 8597, 8605, 8624, 8649, 8668~\AA\
(see Kirkpatrick et al. 1991 and references therein). The strength of
these features increases as the temperature decreases, being more
prominent for giants than for dwarfs. The spectra of late M type stars
are dominated by strong molecular bands showing very weak Ca\,{\sc ii}
lines.

\subsection{Previous index definitions}
\label{previndex}

Several previous works have established different index definitions to
measure the strength of the Ca\,{\sc ii} triplet. In particular, the
most commonly used indices are those defined by JAJ, B\&A, A\&Z, DTT,
A\&D, ZHO, D\&H, and RHS. Most of these indices were defined according
to the {\it classical} definition of line-strength indices, that is,
by establishing a central bandpass covering the spectral feature of
interest, and two other bandpasses (at the red and blue sides of the
central region) which are used to trace a local continuum reference
level through a linear fit to the mean values in both bands (for more
details about the definition and computation of classical indices see
Appendix~A1). In Table~\ref{defs} we list the bandpasses limits of the
previous definitions.

It must be noted that some of these indices slightly differ from the
classical definition. On the one hand, the continuum by ZHO is
computed from the mean value of the 5 highest pixels in each continuum
bandpass, whereas in the index by JAJ, it is chosen relative to the
maximum flux around two selected wavelengths. It is important to
stress that this kind of index definitions is useless for low
signal-to-noise spectra (S/N), leading to a spurious negative
correlation between index value and S/N ratio.  Also, since the
classical index definition refers to the measurement of a unique
spectral feature, previous definitions are given for each Ca\,{\sc ii}
line. In this sense, the most general Ca\,{\sc ii} triplet index is
usually computed as the sum of the three single-line indices, although
some authors (DTT, A\&D and ZHO) prefer to use the sum of the two
strongest lines of the triplet ($\lambda\lambda$~8498.0 and
8542.1~\AA) or even a weighted combination of the three lines (RHS) in
order to optimize the sensitivity of the index to the signal-to-noise
ratio.

\begin{figure*}
\centerline{\hbox{
\psfig{figure=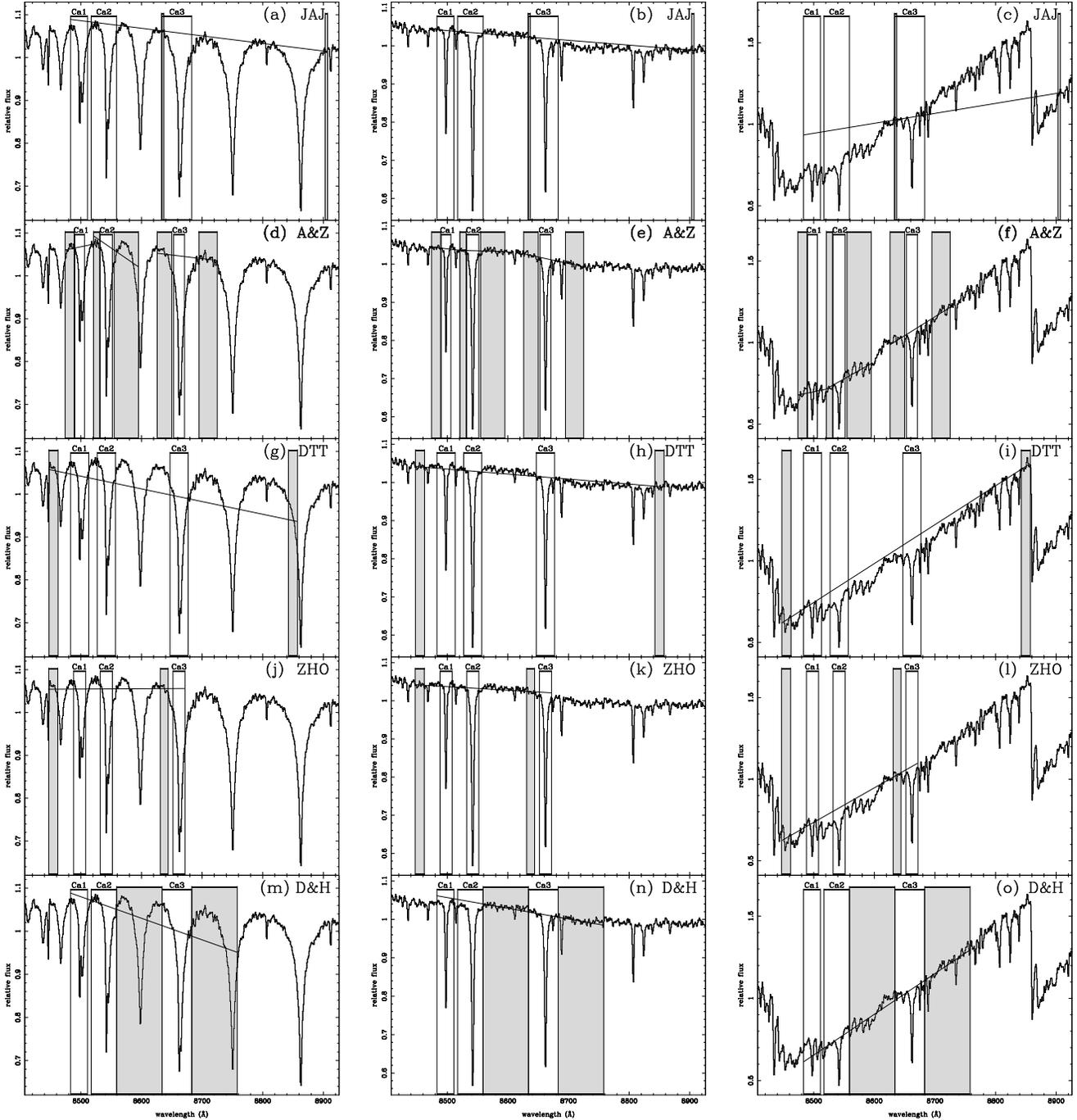}
}}                                                                               
\caption{Previous calcium index definitions over different spectral
types. Indices (from top to bottom) correspond to the systems of JAJ,
A\&Z, DTT, ZHO and D\&H. The representative spectra (from left to
right) are those of HD161817 (A2 VI), HD25329 (K1 Vsb) and HD148783
(M6 III). Grey and open bands mark, respectively, continuum and
central bandpasses, whereas the solid line represents the local
pseudo-continuum computed by an error weighted least-squares fit to
all the pixels in the continuum bands.}
\label{bandasprev}
\end{figure*}

Fig.~\ref{bandasprev} illustrates both the bandpass position and the
predicted continuum of some definitions over three different spectral
types. The indices by B\&A and RHS have not been included since their
blue bandpasses are out of our spectral range. Also the index by A\&D
has been excluded due to its similarity to the index by A\&Z.

\begin{table}
\centering{
\caption{Bandpasses limits of previous calcium indices. Codes are the following: 
JAJ (Jones, Alloin \& Jones, 1984), B\&A (Bica \& Alloin, 1987), A\&Z
(Armandroff
\& Zinn, 1988), DTT (D\'{\i}az, Terlevich \& Terlevich, 1989), A\&D (Armandroff
\& Da Costa, 1991), ZHO (Zhou 1991), D\&H (Delisle \& Hardy, 1992), and RHS
(Rutledge et al., 1997a). Due to the subjective continuum definition
by JAJ, bandpasses limits for this index are as given in D\&H.}
\label{defs}
\begin{tabular}{@{}ccc@{}}
\hline                    
  Index & Central                & Continuum                \\                     
        &         bandpass (\AA) &         bandpasses (\AA) \\ \hline                     
Ca1(JAJ)   & 8483.0--8511.0      &  8633.0--8637.0, 8903.0--8907.0 \\       
Ca2(JAJ)   & 8517.0--8559.0      &  8633.0--8637.0, 8903.0--8907.0 \\ \medskip
Ca3(JAJ)   & 8634.0--8683.0      &  8633.0--8637.0, 8903.0--8907.0 \\ 
Ca1(B\&A)  & 8476.0--8520.0      &  8040.0--8160.0, 8786.0--8844.0 \\       
Ca2(B\&A)  & 8520.0--8564.0      &  8040.0--8160.0, 8786.0--8844.0 \\ \medskip      
Ca3(B\&A)  & 8640.0--8700.0      &  8040.0--8160.0, 8786.0--8844.0 \\ 
Ca1(A\&Z)  & 8490.0--8506.0      &  8474.0--8489.0, 8521.0--8531.0 \\       
Ca2(A\&Z)  & 8532.0--8552.0      &  8521.0--8531.0, 8555.0--8595.0 \\ \medskip      
Ca3(A\&Z)  & 8653.0--8671.0      &  8626.0--8650.0, 8695.0--8725.0 \\ 
Ca1(DTT)   & 8483.0--8513.0      &  8447.5--8462.5, 8842.5--8857.5 \\       
Ca2(DTT)   & 8527.0--8557.0      &  8447.5--8462.5, 8842.5--8857.5 \\ \medskip      
Ca3(DTT)   & 8647.0--8677.0      &  8447.5--8462.5, 8842.5--8857.5 \\ 
Ca1(ZHO)   & 8488.0--8508.0      &  8447.0--8462.0, 8631.0--8644.0 \\ 
Ca2(ZHO)   & 8532.0--8552.0      &  8447.0--8462.0, 8631.0--8644.0 \\ \medskip
Ca3(ZHO)   & 8652.0--8672.0      &  8447.0--8462.0, 8631.0--8644.0 \\ 
Ca2(A\&D)  & 8532.0--8552.0      &  8474.0--8489.0, 8559.0--8595.0 \\ \medskip
Ca3(A\&D)  & 8653.0--8671.0      &  8626.0--8647.0, 8695.0--8754.0 \\ 
Ca1(D\&H)  & 8483.0--8511.0      &  8559.0--8634.0, 8683.0--8758.0 \\       
Ca2(D\&H)  & 8517.0--8559.0      &  8559.0--8634.0, 8683.0--8758.0 \\ \medskip     
Ca3(D\&H)  & 8634.0--8683.0      &  8559.0--8634.0, 8683.0--8758.0 \\ 
Ca1(RHS)   & 8490.0--8506.0      &  8346.0--8489.0, 8563.0--8642.0 \\       
Ca2(RHS)   & 8532.0--8552.0      &  8346.0--8489.0, 8563.0--8642.0 \\      
Ca3(RHS)   & 8653.0--8671.0      &  8563.0--8642.0, 8697.0--8754.0 \\ 
\hline
\end{tabular}
}
\end{table}

In order to predict a reliable local continuum, the continuum
bandpasses should be located in spectral regions free from strong
absorption features. In this sense, all the previous definitions are
optimized for G and K spectral types (see Figs.~\ref{bandasprev}b,
\ref{bandasprev}e, \ref{bandasprev}h, \ref{bandasprev}k and
\ref{bandasprev}n). However, because the relative contribution of the
earliest spectral types to the integrated spectrum of stellar systems
was expected to be negligible at this spectral range, most of the
previous calcium indices were defined without taking into account the
wavelength position of the Paschen series. The indices by A\&Z, DTT,
and D\&H clearly suffer from such a limitation deriving a continuum
level below the true one for hot stars (see Figs.~\ref{bandasprev}d,
\ref{bandasprev}g, and \ref{bandasprev}m).
On the other hand, because of the presence of molecular bands in late
type stars, some definitions show unreliable continuum levels for
these spectral types. In particular, the red bandpass of the index by
JAJ falls in a strong TiO absorption (Fig.~\ref{bandasprev}c), while
the blue bandpasses by DTT and ZHO (which are roughly the same) are
located too close to the continuum break, causing a rise in the
continuum level at the position of the Ca lines (see
Figs.~\ref{bandasprev}i and \ref{bandasprev}l). Also, when the
relative distance between the two continuum bandpasses is too large,
the index can be insensitive to local variations in the true continuum
(JAJ and DTT), whereas indices with continuum sidebands which are too
close (A\&Z) are highly sensitive to velocity dispersion broadening
(see Section~\ref{veldisp}).  As far as the width of the sidebands is
concerned, bandpasses that are too wide (D\&H) are not appropriate
when the spectral region is full of absorption lines which decrease
the continuum level. Indices with two narrow sidebands (JAJ) are very
sensitive to the S/N ratio and to residuals from sky
subtraction. Another important factor is the width of the central
bands, since, as we will see in Section~\ref{sensitivity}, it mainly
determines the sensitivity of the index to the S/N ratio and the
velocity dispersion broadening.  Finally, and not the least important,
it is obvious that any Ca index definition following the classical
system is unavoidably affected by Paschen contamination in the central
bandpass for stars hotter than G1 and G3, for dwarfs and giants
respectively.

\subsection{New indices definitions}
\label{newdefs}

Although the previous index definitions have been very useful to
increase the understanding of the behaviour of the Ca\,{\sc ii}
triplet both in individual stars and stellar populations (see
Section~\ref{prevref}), in this work we have decided to define a new
set of improved indices, specially designed to be measured in the
integrated spectra of galaxies. With the new definitions we have tried
to alleviate difficulties such as the continuum definition, the
Paschen contamination, and the sensitivity to the S/N ratio and the
velocity dispersion (spectral resolution).  Note that the aim of these
new indices is not to provide measurements as close as possible to the
true equivalent widths. Other indices, like those of RHS are much more
appropriate for that purpose.

The new indices have been defined according to a new type of
line-strength index concept: the {\it generic} index. It is a natural
generalization of the classical definition which includes the
following requirements: it is characterized by an arbitrary number of
continuum and spectral-feature bandpasses, the contribution of each
spectral-feature bandpass can be modified by defining a multiplicative
factor for each one, and the pseudo-continuum is derived by using an
error weighted least-squares fit to all the pixels of the continuum
bandpasses. Although at first sight it could seem that this approach
is exactly the same that the one obtained by adding several classical
indices (weighted with the corresponding multiplicative factors), the
fact that all the spectral features share the same continuum has an
important effect in the computation of the index error as it is
explained in Appendix~A2. It is also convenient to highlight that the
requirements incorporated in the definition of generic indices
translate into important improvements:

\begin{enumerate}
\item Generic indices are specially suited for the measurement of
adjacent spectral features, where the use of a common continuum level
can be useful. They are also highly recommended when the spectral
region of interest is densely populated by spectral features, since
many thin continuum bandpasses at adequate locations make it possible
to avoid the presence of the other spectral features. Moreover, a
large number of continuum bands guarantees a low sensitivity to the
S/N ratio and a robust continuum definition.
\item As we will prove later, the multiplicative factors are
useful to remove the contamination by other spectral features
(either in absorption or in emission).
\item The use of an error weighted least-squares fit in the
determination of the local pseudo-continuum is specially advantageous
when we measure the near-IR absorption features. In this spectral
region, the presence of sky emission lines and telluric absorptions
implies that the signal-to-noise ratio, as a function of wavelength,
is a highly inhomogeneous function. 
\end{enumerate}

Following the notation from Cohen (1978) for the calcium triplet, the
new generic index will be referred as CaT. It is defined by
establishing five continuum bandpasses, and three central bandpasses
covering each calcium line (Ca1, Ca2 and Ca3). The continuum regions
were carefully chosen to optimize the continuum level for all the
spectral types, even when the spectra were broadened up to $\sigma =
300$~km~s$^{-1}$. In order to calibrate the CaT contamination by the
Paschen series in early spectral types, we defined a new generic
index, namely PaT, which measures the strength of three Paschen lines
free from the calcium contamination. The continuum bandpasses are the
same as in CaT whereas the spectral-feature bandpasses (Pa1, Pa2 and
Pa3) are centered on the lines P17, P14 and P12 of the series. In both
definitions, the multiplicative factors equal unity and the analytical
expressions are

\begin{equation} 
{\rm CaT} = {\rm Ca1} + {\rm Ca2} + {\rm Ca3}, \;\; \; {\rm and}
\end{equation}
\begin{equation} 
{\rm PaT} = {\rm Pa1} + {\rm Pa2} + {\rm Pa3}.   
\end{equation}

\begin{figure}
\centerline{\hbox{
\psfig{figure=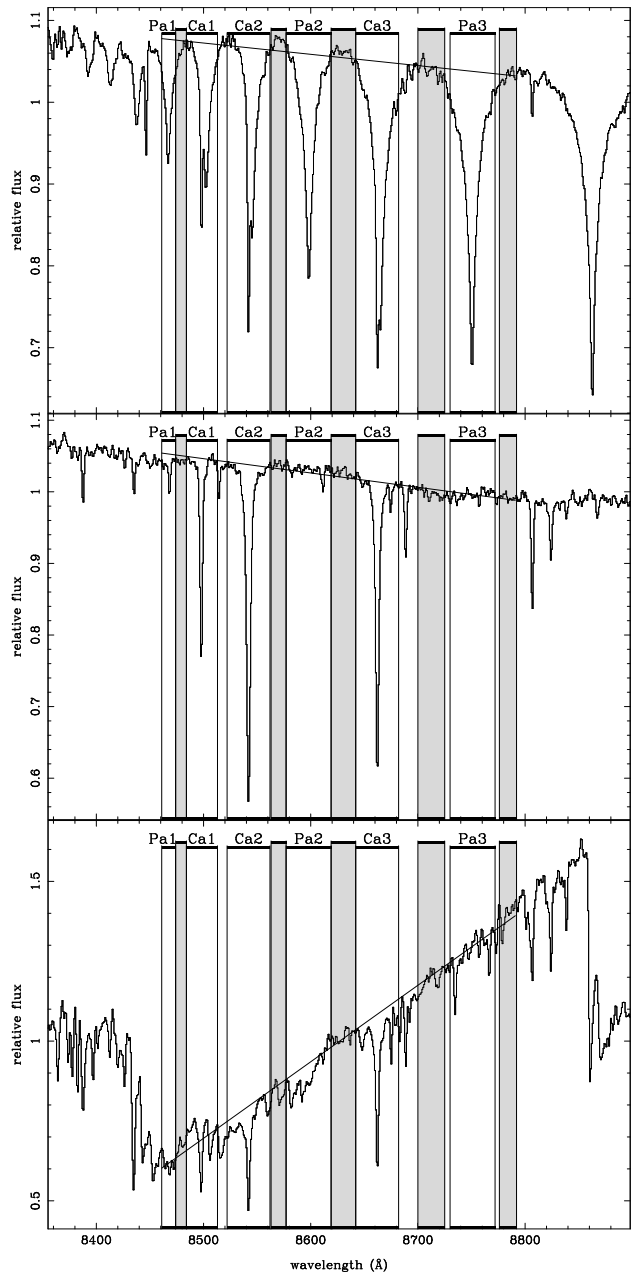}
}}
\caption{CaT and PaT indices over different spectral types. 
The spectra correspond (from top to bottom) to HD161817 (A2 VI),
HD25329 (K1 Vsb) and HD148783 (M6 III), which are the same spectral
types displayed in Fig.~\ref{bandasprev}. Grey and open bands
represent continuum and central bandpasses respectively, with the
solid line showing the derived local pseudo-continuum.}
\label{bandasCaTPaT}
\end{figure}

The bandpasses limits for the CaT and PaT indices are listed in
Table~\ref{defCaTPaT}, and the bandpasses positions and predicted
continua are illustrated in Fig.~\ref{bandasCaTPaT} for different
spectral types.

\begin{table}
\centering{
\caption{Bandpasses limits for the generic indices CaT and PaT.}
\label{defCaTPaT}
\begin{tabular}{@{}ccc@{}}
\hline                    
CaT central        & PaT central        & Continuum        \\                     
bandpasses (\AA)   & bandpasses (\AA)   & bandpasses (\AA) \\ 
\hline                     
Ca1 8484.0--8513.0 & Pa1 8461.0--8474.0 & 8474.0--8484.0   \\       
Ca2 8522.0--8562.0 & Pa2 8577.0--8619.0 & 8563.0--8577.0   \\       
Ca3 8642.0--8682.0 & Pa3 8730.0--8772.0 & 8619.0--8642.0   \\
                   &                    & 8700.0--8725.0   \\
                   &                    & 8776.0--8792.0   \\ 
\hline
\end{tabular}
}
\end{table}

It is important to note that, apart from rough corrections applied to
the integrated spectra of young star clusters (Alloin \& Bica 1989)
and theoretical considerations on synthetic spectra (Chmielewski
2000), few previous works have faced the Paschen contamination in
individual stars.

Finally, we have defined a new calcium triplet index, namely
CaT$^{*}$, which expresses the strength of the Ca\,{\sc ii} lines
corrected from the contamination by Paschen lines. The new index was
defined by imposing the following requirements. First, the values of
CaT$^{*}$ should be very similar to those of CaT for late-type
stars. Secondly, since the true calcium strength in hot stars is very
low or even null, the CaT$^{*}$ index should tend to values around
zero for the earliest spectral types. Note that, since the CaT index
in hot stars is actually measuring the strength of the Paschen lines
rather than those of Ca, such an index (as well as those used in
previous works) can lead to wrong conclusions when interpreting the
integrated spectra of young stellar populations.

\begin{figure}
\centerline{\hbox{
\psfig{figure=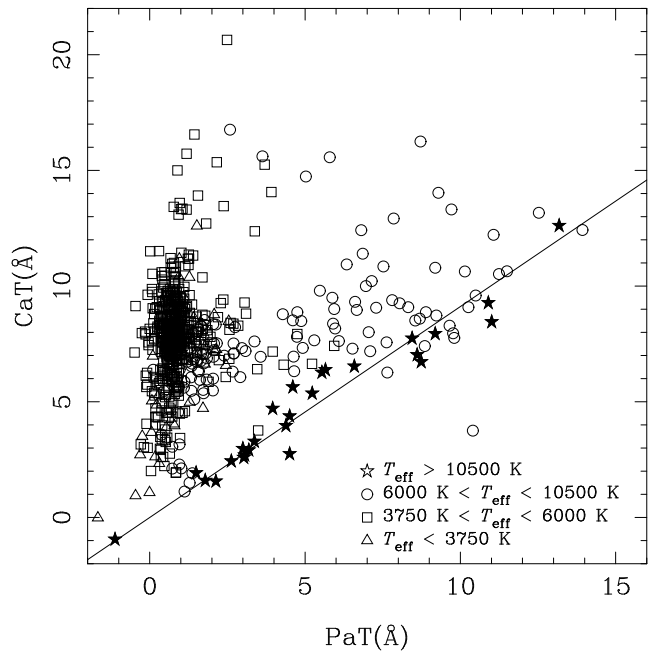}
}} 
\caption{CaT -- PaT diagram for the whole stellar library. Different symbols
are used to indicate different ranges of effective temperature, as
shown in the key. The solid line CaT$\;=0.93\;$PaT represents an error
weighted least-squares fit to the 26 filled symbols.}
\label{CaTPaTfit}
\end{figure}

To estimate the hydrogen contribution to the calcium index, we have
compared the CaT and PaT indices for hot stars, which show a pure
smooth Paschen series in their spectra.  For this subsample, we should
expect a linear relationship between the CaT index (which measures the
P16, P15 and P13 lines falling in the Ca1, Ca2 and Ca3 bandpasses) and
the PaT index (measuring the P17, P14 and P12 lines). Such a relation
was calibrated by using an error weighted least-squares fit to the 26
library stars with {\it T}$_{\rm eff}$ $>$ 10500~K (see
Fig.~\ref{CaTPaTfit}) obtaining
\begin{equation} 
\label{slopeCaTPaT}
{\rm CaT} = (0.93 \pm 0.03)\;{\rm PaT}.
\end{equation}
Thus the Paschen corrected CaT$^{*}$ index
was defined by subtracting from CaT the correction term given
by the above equation, i.e.
\begin{equation} 
{\rm CaT}^{*} = {\rm CaT} - 0.93\;{\rm PaT},
\end{equation}
which can be considered, in turn, as a generic index with five
continuum bands, six central bandpasses and two different
multiplicative factors (1.00 and $-$0.93). Since we are adopting a
definition, the multiplicative factor $-$0.93 is assumed to be exact,
so we are not concerned with the nominal error 0.03 obtained in the
fit to the hot stars.

Finally, it is worth noting that CaT$^{*}$ is also useful for late
type stars where PaT takes values around zero and the correction term
becomes negligible.

For those readers interested in measuring the new indices in their
spectra, we provide a set of public {\sc fortran} subroutines
available from the website given at the end of the paper.

\subsection{Sensitivities of the indices to different effects}
\label{sensitivity}

In this section, we discuss the sensitivity of the new and previous Ca
indices to the signal-to-noise ratio, velocity dispersion broadening
(or spectral resolution), relative flux calibration and sky
subtraction.

\subsubsection{Signal-to-noise ratio}
\label{snsens}

In Appendix~A we show in detail how to compute random errors, arising
from photon statistics, for the new generic indices. Following a
procedure similar to that presented in Cardiel et al. (1998, hereafter
CGCG), we have derived accurate formulae for the computation of random
errors for the new Ca indices (Appendix~A2). Note that reliable random
errors can only be derived when, after a full control of the error
propagation throughout the data reduction, an accurate error spectrum
is derived for each data spectrum.  Appendix~A3 also provides
analytical estimates of the predicted random errors as a function of
signal-to-noise (S/N) ratio per angstrom of the form
\begin{equation}
\sigma[{\cal I}_{\rm a}] \simeq
  \displaystyle \frac{c_1-c_2\;{\cal I}_{\rm a}}{S\!N(\rm\AA)}.
\label{errorvssn}
\end{equation}
The exact values of the $c_i$ coefficients are given in
Eqs.~(\ref{relaciones_aproximadas_errores}). As an illustration, for
an M0 giant, a S/N per angstrom of 16 is required to measure the CaT
and CaT$^{*}$ indices with a 10 per cent uncertainty. We have selected
this spectral type since, as we will see in Paper~IV, old stellar
populations exhibit near-IR spectra close to those or early-M stars.

To compare the sensitivity of our indices to the S/N ratio with that
of previous definitions, we have computed the typical random error due
to photon noise ($\sigma[{\cal I}_{\rm a}]$) for the 706 spectra of
the library and all the considered index definitions.
Table~\ref{SNsensit} lists the ratio between the relative random error
of any index and that of CaT. The relations were computed through a
least-squares linear fit (rejecting data outside the 99.73 per cent
confidence level) to the relative errors of the whole sample.  It can
be seen that our index is similar to that of D\&H, but leads to
somewhat larger random errors than the definitions of ZHO, A\&Z and
DTT.  The reason for this is that these systems use relative narrow
central bandpasses (with a total width for the Ca2 and Ca3 lines of
38, 40 and 60~\AA\ respectively, compared to the 80~\AA\ in our
definition). Note that since the noise in the bottom of any absorption
line is lower than in the wings, indices with narrow central
bandpasses yield small random errors.  Unfortunately, this turns into
a high sensitivity to velocity dispersion broadening (as we will see
in Section~\ref{veldisp}). On the other hand, the index from JAJ uses
a wide central band and extremely narrow continuum bandpasses which
increase the uncertainties in the determination of the continuum
level. This is why this index shows a high sensitivity to the
signal-to-noise ratio.

\begin{table}                                                              
\centering{
\caption{Ratio between the relative random error of any index and that of CaT.}
\label{SNsensit}                                                            
\begin{tabular}{cc}
\hline
${\cal I}_{\rm a}$ & $\displaystyle\frac{\sigma[{\cal I}_{\rm
a}]/{\cal I}_{\rm a}}{\sigma[{\rm CaT}]/{\rm CaT}}$\\
\hline
CaT(ZHO)        & $0.788 \pm 0.004$ \\
CaT(D\&H)       & $1.009 \pm 0.006$ \\
CaT(JAJ)        & $3.938 \pm 0.046$ \\
CaT(DTT)        & $0.933 \pm 0.007$ \\
CaT(A\&Z)       & $0.567 \pm 0.002$ \\
\hline
\end{tabular}
}
\end{table}

As it was also noted in Section~\ref{previndex}, some Ca indices have
been defined as the sum of the two strongest Ca\,{\sc ii} lines in
order to increase the S/N ratio of the final measurement. In our case,
when the Ca1 line is removed from the CaT definition, the relative
random errors decrease by a factor of 10 per cent.  However, keeping in
mind that the central bandpass of the Ca3 line falls near the region
of strong telluric absorptions, being completely immersed in it for
galaxies with $z>0.03$, we have decided to include the bluest line of
the triplet (Ca1) in the index definition, at the expense of
sacrificing the gain in accuracy, to ensure a more reliable index
measurement for galaxies with relatively high radial velocities.

\subsubsection{Spectral resolution and velocity dispersion broadening}
\label{veldisp}

In order to study the sensitivity of the Ca indices to the spectral
resolution or velocity dispersion ($\sigma$), we selected a subsample
of near-solar metallicity templates of all the spectral types and
luminosity classes and broadened their spectra with additional
$\sigma$'s from $\sigma = 25$~km~s$^{-1}$ up to $\sigma =
400$~km~s$^{-1}$ (in steps of 25~km~s$^{-1}$). The indices were
measured in the full set of broadened spectra, and we computed, for
each group of templates with the same spectral type, a third-order
polynomial fit to the relative changes of the index values as a
function of velocity dispersion
\begin{equation} 
{\frac{{\cal I}_{\rm a}(\sigma) - {\cal I}_{\rm a}
(\sigma_{0})}{{\cal I}_{\rm a}(\sigma)} 
= a + b\;\sigma + c\;\sigma^{2} + d\;\sigma^{3} \equiv  p(\sigma)},
\label{poly}
\end{equation}
where $\sigma_{0} = 22.2$~km~s$^{-1}$, corresponding to the final
spectral resolution of the stellar library (FWHM = 1.50~\AA, see
Section~\ref{systematic}). For a given spectral type, we did not
compute separate fits for the different luminosity classes since the
derived polynomials were very similar. Using the above equation, it is
easy to show that the indices measured at two different velocity
dispersions, $\sigma_1$ and $\sigma_2$, are related by
\begin{equation} 
{\cal I}_{\rm a}(\sigma_{2}) = {\cal I}_{\rm a}(\sigma_{1})\frac{1 - p(\sigma_{1})}{1 - p(\sigma_{2})}.
\label{relpoly}
\end{equation}

\begin{figure*}
\centerline{\hbox{
\psfig{figure=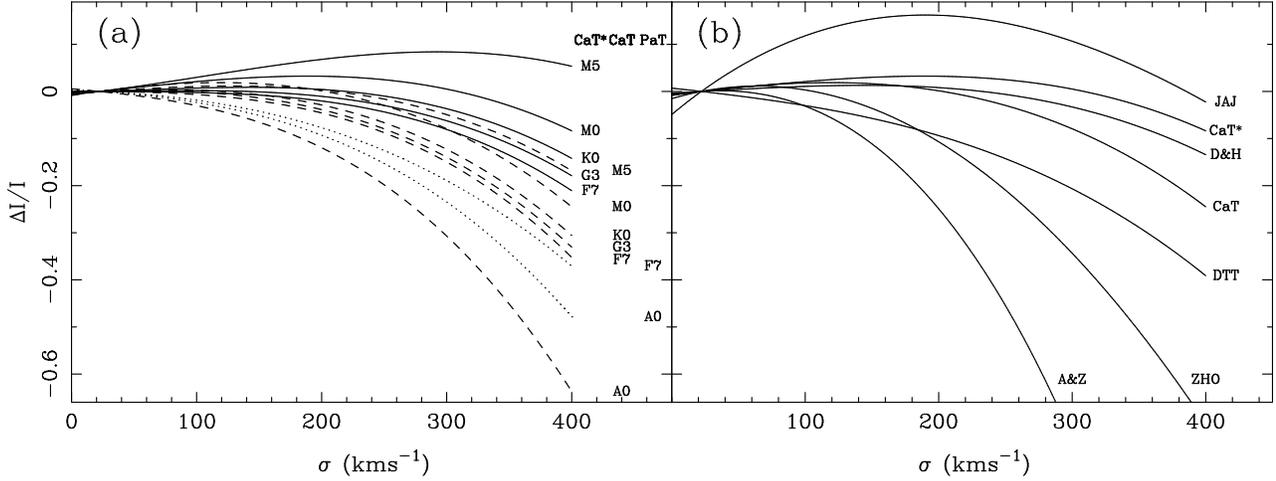}
}} 
\caption{Index sensitivity to the spectral resolution or velocity
dispersion broadening. Fig.~\ref{dispsens}a shows the relative
variation of CaT$^{*}$ (solid lines), CaT (dashed lines) and PaT
(dotted lines) for several broadened spectral types. In
Fig.~\ref{dispsens}b, similar polynomials are shown for the new and
the previous Ca\,{\sc ii} indices for an M0 type star. In both
diagrams, $\Delta I/I$ is zero for $\sigma = 22.2$~km~s$^{-1}$ (see
Eq.~(\ref{poly})), corresponding to the spectral resolution of the
stellar library (FWHM = 1.50~\AA).}
\label{dispsens}
\end{figure*}

Table~\ref{polcoef} lists the derived coefficients for CaT$^{*}$, CaT
and PaT measured over different spectral types. Note that we only
provide calibrations for spectral types with equivalent widths
significantly different from zero (e.g. stars cooler and hotter than
mid-F types for CaT$^{*}$ and PaT respectively). For illustration,
such polynomials are shown in Fig.~\ref{dispsens}a. It is readily
observed that the sensitivity of CaT$^{*}$ to the velocity dispersion
broadening is much lower than that of CaT and PaT. Such an improvement
is explained since the effects of the broadening on CaT and PaT are
partially compensated when the CaT$^{*}$ index is computed. This fact
should be considered as another additional advantage of employing
generic indices for the measurement of spectral features.

\begin{table}                                                              
\centering{
\caption{Coefficients of the broadening correction polynomials ($\Delta
{\cal I}_{\rm a}/{\cal I}_{\rm a}$ = $a$ + $b\;\sigma$ +
$c\;\sigma^{2}$ + $d\;\sigma^{3}$) for the new indices and different
spectral types.}
\label{polcoef}                                                            
\begin{tabular}{@{}ccr@{}lr@{}lr@{}lc@{}}
\hline
 ${\cal I}_{\rm a}$ & SpT& \multicolumn{2}{c}{$a$($\times$10$^{-3}$)} & 
\multicolumn{2}{c}{$b$($\times$10$^{-5}$)} & 
\multicolumn{2}{c}{$c$($\times$10$^{-7}$)} & $d$($\times$10$^{-9}$)\\
\hline     
CaT*&F7&  \ 0.&784 & --4.&683 &   6.&197 &--4.563\\
    &G3&\ --0.&452 &   1.&048 &   5.&386 &--4.201\\
    &K0&\ --1.&565 &   5.&605 &   7.&483 &--4.418\\
    &M0&\ --6.&240 &  27.&524 &   3.&468 &--3.797\\
    &M5&\ --6.&795 &  27.&980 &  12.&715 &--3.986\\
&&&&&&&&\\
CaT &A0&  \ 5.&562 &--23.&662 & --4.&621 &--7.408\\
    &F7&  \ 0.&503 & --2.&119 &   0.&559 &--5.523\\
    &G3&\ --1.&969 &  10.&305 & --5.&495 &--4.407\\
    &K0&\ --3.&106 &  15.&413 & --5.&453 &--4.332\\
    &M0&\ --8.&542 &  41.&116 &--11.&118 &--3.485\\
    &M5&\ --3.&953 &  17.&938 &   0.&243 &--3.751\\
&&&&&&&&\\
PaT &A0&  \ 3.&943 &--16.&146 & --6.&162 &--4.991\\
    &F7&  \ 2.&169 & --7.&226 &--10.&861 &--2.664\\
\hline
\end{tabular}                                                              
}                                                                          
\end{table}

The measurements and fitting procedures given above were also
performed for all the previous Ca\,{\sc ii} indices, although only the
calibrations computed for M0 type spectra (see Section~\ref{snsens})
are presented in this paper. The derived coefficients are presented in
Table~\ref{polcoef2}, whilst the polynomials, together with those of
CaT and CaT$^*$ are shown in Fig.~\ref{dispsens}b. It is clear that,
except for the index by D\&H, all the previous indices are more
sensitive to broadening than our definitions. In most cases, the high
sensitivity is due to the relatively narrow central bandpasses (A\&Z,
ZHO and DTT). The index by JAJ has wide central bandpasses, but the
blue continuum band is located too close to the Ca3 line. Finally, the
index by D\&H, with wide continuum and central bandpasses, is the only
one which overcomes the sensitivity to broadening of CaT, with a
similar behaviour to our CaT$^{*}$ index.

\begin{table}
\centering{
\caption{Coefficients of the broadening correction polynomials
($\Delta {\cal I}_{\rm a}/{\cal I}_{\rm a}$ = $a$ + $b\;\sigma$ +
$c\;\sigma^{2}$ + $d\;\sigma^{3}$) for previous index definitions and an
M0 spectral type.}
\label{polcoef2}                                                            
\begin{tabular}{@{}r@{}lr@{}lr@{}lr@{}lr@{}l@{}}
\hline
\multicolumn{2}{c}{${\cal I}_{\rm a}$} & \multicolumn{2}{c}{$a$($\times$10$^{-3}$)} & \multicolumn{2}{c}{$b$($\times$10$^{-5}$)} & \multicolumn{2}{c}{$c$($\times$10$^{-7}$)} & \multicolumn{2}{c}{$d$($\times$10$^{-9}$)}\\
\hline     
CaT&(A\&Z) & --7.&712 &   48.&585 & --59.&578 & --12.&551 \\
CaT&(D\&H) & --3.&517 &   15.&486 &    2.&396 &  --3.&610 \\
CaT&(DTT)  &   7.&107 & --31.&378 &  --2.&028 &  --3.&755 \\
CaT&(JAJ)  &--49.&068 &  237.&304 & --74.&233 &    4.&143 \\
CaT&(ZHO)  &--15.&283 &   83.&299 & --65.&193 &    0.&314 \\
\hline
\end{tabular}                                                              
}                                                                          
\end{table}

\subsubsection{Flux calibration}
\label{fluxcalx}

There is no simple recipe to estimate the sensitivity of an index to
the uncertainties in the relative flux calibration, since it not only
depends on the index definition but also on the response curve of the
instrumental configuration. In general, the effect is larger as the
wavelength coverage of the sidebands increases. As an illustration,
for the JKT runs, the systematic differences between the CaT index
measured after and before the flux calibration were around $\Delta$CaT
= 0.4~\AA.  This value is higher than typical uncertainties due to
random errors and any other systematic effects (as we will see in
Section~\ref{errors}), and stresses the importance of an accurate flux
calibration before performing any meaningful comparison between
evolutionary synthesis model predictions and measured spectra.

\subsubsection{Sky subtraction}
\label{subsky}

One of the major problems to face in the measurement of spectral
features in the near-IR is the presence of both, emission lines
produced by the OH radical (e.g. Rousselot et al. 2000), and telluric
absorptions arising from water vapour and other molecules (e.g. Stevenson
1994, Chmielewski 2000). A proper removal of these effects heavily
relies on the absence of systematic errors in the data reduction. For
instance, biases in wavelength calibration (even small sampling
aliasing) and in flux calibration (improper flatfielding, residual
fringing, and variable slit width along the slit, among others)
seriously compromise an appropriate sky subtraction. These problems
can be specially severe when observing faint sources, where the
systematic deviations produced by an improper sky subtraction likely
exceed random errors. Since the actual relevance of the mentioned
uncertainties is a function of many variables, including the radial
velocity of the targets, a case by case analysis must be performed to
cope with them. Note, however, that the availability of error frames
(in which pixels at the position of sky emission lines have larger
errors) together with the use of error weighted least-squares fits to
determine the pseudo-continuum (see Section~\ref{newdefs}) alleviate
the effects of sky residuals in the measurement of generic indices.

\subsection{Conversions between different calcium index systems}
\label{calibrations}

\begin{table*}
\centering{
\caption{Calibrations between different calcium index systems. CaT:
New calcium index defined in this paper. CaT({\it i}): Calcium index
in the system by {\it i} measured in our spectra. CaT({\it i})$_{j}$: Calcium
index in the system by {\it i} measured in the spectra by {\it j}. Codes: JAJ
(Jones et al. 1984), A\&Z (Armandroff \& Zinn 1988), DTT (D\'{\i}az et
al. 1989), ZHO (Zhou 1991), D\&H (Delisle \& Hardy 1992) and ITD
(Idiart, Thevenin \& de Freitas Pacheco 1997). $\sigma$:
unbiased standard deviation of the fit. N: Number of stars. Range:
Effective temperature region where the calibration was obtained.}
\label{compstab}
\begin{tabular}{@{}r@{}r@{}lccc@{}}
\hline
\multicolumn{3}{c}{Calibrations}& $\sigma$ & $N$ & {\it T}$_{{\rm eff}}$ range (K)\\
\hline
CaT       =&   2.064 + &\ 0.738 CaT(JAJ)                     & 0.86 & 627 & 3600--38400\\
CaT       =& --0.155 + &\ 1.183 CaT(A\&Z)                    & 0.34 & 582 & 3600--7500 \\ 
CaT       =&   1.450 + &\ 1.000 CaT(DTT)                     & 0.39 & 486 & 4000--6300 \\ 
CaT       =&   0.002 + &\ 1.345 CaT(ZHO)                     & 0.35 & 591 & 3000--7500 \\ \medskip
CaT       =& --0.756 + &\ 1.107 CaT(D\&H)                    & 0.36 & 559 & 3600--7000 \\ 
CaT(JAJ)  =& --3.322 + &\ 1.524 CaT(JAJ)$_{{\rm JAJ}}$       & 1.15 &\ 21 & 3600--6500 \\
CaT(A\&Z) =&   0.000 + &\ 1.000 CaT(A\&Z)$_{{\rm ITD}}$      & 0.23 &\ 41 & 4350--6300 \\
CaT(DTT)  =& --0.917 + &\ 1.103 CaT(DTT)$_{{\rm DTT}}$       & 0.55 & 102 & 3425--6800 \\  
CaT(ZHO)  =&   1.746 + &\ 0.616 CaT(ZHO)$_{{\rm ZHO}}$       & 0.66 &\ 36 & 3350--6250 \\
\hline
\end{tabular}
}
\end{table*}

\begin{figure*}
\centerline{\hbox{
\psfig{figure=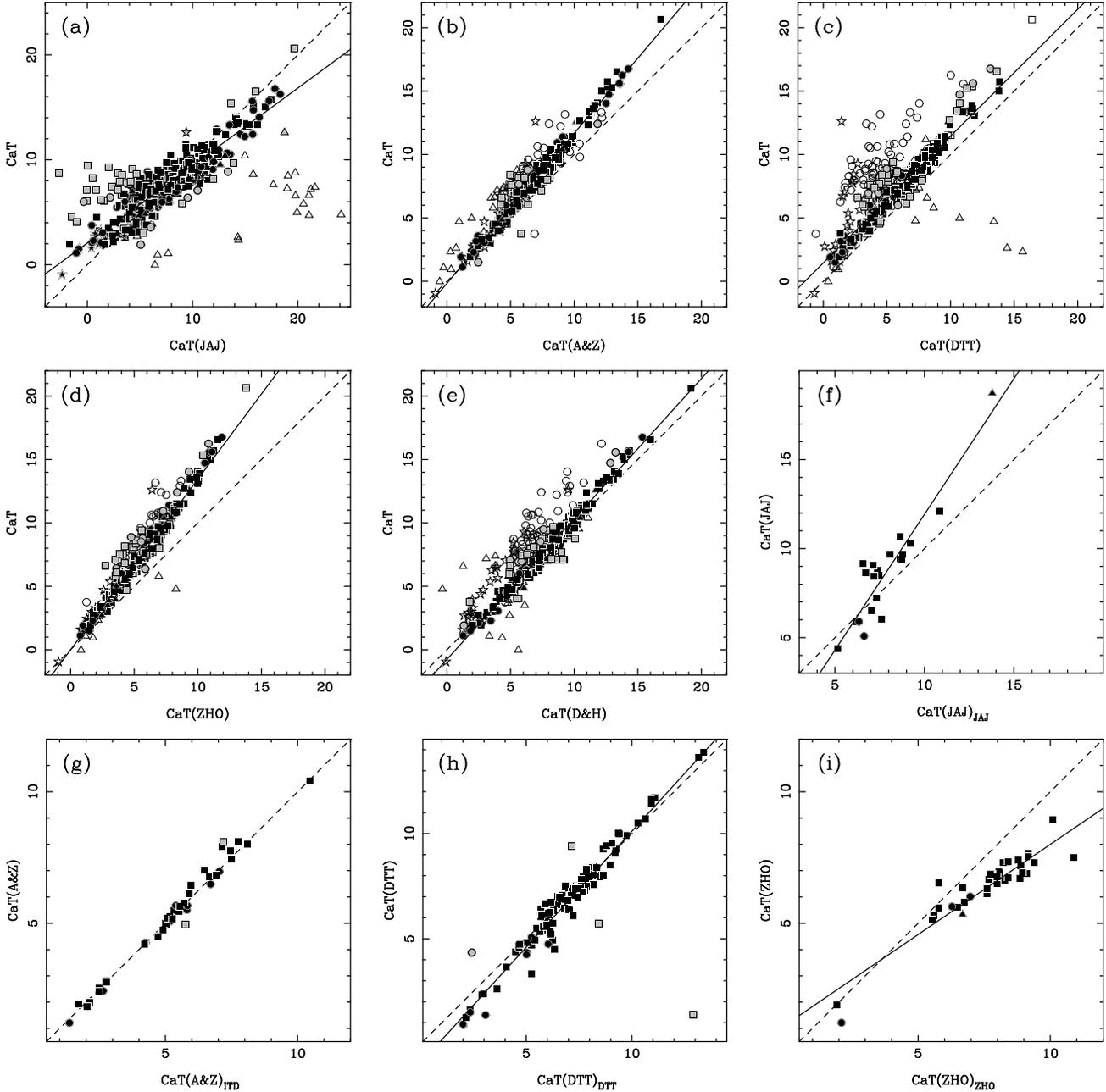}
}}
\caption{Comparison between different systems. Diagrams (a) to (e)
compare the CaT index with the previous indices measured over the 706
stars of the stellar library. CaT($i$) means the index in the system by $i$ measured
over our spectra. Diagrams (f) to (i) compare measurements of the same
index in our and other spectra for a subsample of stars in common.
CaT($i$)$_{j}$ means the index in the system by $i$ as measured by $j$.
Label codes are JAJ (Jones et al. 1984), A\&Z (Armandroff \& Zinn
1988), DTT (D\'{\i}az et al. 1989), ZHO (Zhou 1991), D\&H (Delisle \&
Hardy 1992) and ITD (Idiart, Thevenin \& de Freitas Pacheco
1997). Symbol types, indicating different ranges of effective
temperature, are the same as in Fig.~\ref{CaTPaTfit}. The dashed line
shows the one-to-one relation. Grey symbols are stars deviating
more than $3\sigma$ from the fitted relation, whereas open
symbols refer to those stars with effective temperatures outside the fitted range.
The solid line marks the most significant fit to
the symbols in black. In all the cases, indices are measured in angstroms}
\label{compsplot}
\end{figure*}

In this section we give a set of calibrations to make conversions
between the different systems of calcium indices. These will be useful
for readers interested in using our system or in transforming their
own calcium index data into the new indices.

First, we have compared the measurements of the previous calcium index
definitions over the 706 library stars with the corresponding CaT
values. The calibrations were computed by deriving an error weighted
least-squares fit to a straight line. If the slope of such fit was not
significantly different from one, we fitted a constant offset. If this
term was not significant, we kept the one-to-one relation. Due to
the locations of the sidebands of some previous definitions (see
Section~\ref{previndex}), hot and cold stars may depart from the
general trend of the rest of the library stars. When this occurs, and
in order to ensure the quality of the fit, we have restricted the
range of the calibration excluding from the fit those stars with
extreme temperatures. In particular, hot stars have not been used for
the comparison with A\&Z, DTT, ZHO and D\&H, where the indices were
underestimated due to low continuum levels.  Similarly, the high
continuum levels derived for the cold stars in the systems by JAJ and
DTT lead to unreliable indices. Moreover, stars within the $T_{\rm
eff}$ range but deviating more than $3\sigma$ from the fitted relation
were also rejected. Most of them are cluster stars with low S/N ratios
which present high uncertainties in their Ca measurements. This
becomes specially significant in the comparison with JAJ, whose index
degrades rapidly as the S/N ratio decreases, producing the large
dispersion in the fitted relation.  Figs.~\ref{compsplot}a to
\ref{compsplot}e show the calibrations explained above, including the
fitted and rejected points and the final fit. CaT($i$) refers to the
index defined by the reference $i$ measured on our spectra.

Further calibrations are also presented for those papers which measure
a calcium index over a sample of stars in common with our stellar
library. Following the same procedure to compute the fit, the
published index values for the stars in common were compared with the
ones obtained by measuring their index over the same stars in our
stellar library. It must be noted that, since the measured index is
the same in these comparisons, systematic differences between the two
sets of measurements are mainly due to flux calibration effects (some
authors did not perform any flux calibration) and differences in
spectral resolution (although, given the previous resolutions, see
Table~\ref{tab_lib}, this is not the dominant effect).  The importance
of such effects are shown in Figs.~\ref{compsplot}f to
\ref{compsplot}i, where CaT($i$)$_{j}$ now refers to the index defined by $i$
and measured by $j$.

In Table~\ref{compstab} we present the derived calibrations together
with other details of the fits. These fits can be useful for those
readers who want to convert their data of previous Ca indices into the
new CaT system. The upper calibrations in Table~\ref{compstab} should
only be applied once the spectra are on the same spectrophotometric
system (i.e. equally flux calibrated) and spectral resolution as the
sample of this work. If that is not the case, calibrations like the
lower ones in Table~\ref{compstab} should be applied before.

Readers interested in using our Ca system to make comparison of
observed line-strengths with the fitting-function predictions that
will be presented in the upcoming papers of this series, are highly
recommend to convert their observations to this system by:
\begin{enumerate}
\item flux calibrating the spectra,
\item observing a significant sample of stars in common with our stellar
library,
\item converting to our spectral resolution using the relations given in
Section~\ref{veldisp} and those that will appear in Paper~IV for the
integrated spectra of galaxies.
\end{enumerate}

\section{Computation of the index errors}
\label{errors}

One of the principal objectives of this series is to obtain a
functional representation of the behaviour of the calcium triplet as a
function of the atmospheric stellar parameters (see Paper~III). In
order to ensure the precision of the final fitting functions, an
accurate treatment of the errors is needed. To do this, two kind of
error sources must be considered: the uncertainties in the atmospheric
parameters (which are thoroughly discussed in Paper~II) and those
associated with the index measurements, which are the subject of this
section.

Following the error analysis described in Gorgas et al. (1999), we
divide the error sources into random errors and systematic effects.
In brief, the accuracy of the derived random errors will be checked by
the comparison of stars with repeated observations within the same
run, whereas the systematic errors will be mainly derived by
comparisons of stars in common between different runs.

\subsection{Random errors}
\label{random}

We consider three principal sources of random errors: photon
statistics, flux calibration and the combined effect of wavelength
calibration and radial velocity uncertainties. Therefore, expected
random errors for the $i^{{\rm th}}$ star can be computed through the
following quadratic addition
\[ \sigma^{2} [{\cal I_{\rm a}}]_{{\rm expected},i} = \]
\begin{equation} 
= \sigma^{2}[{\cal I_{\rm a}}]_{{\rm photon},i} +
\sigma^{2}[{\cal I_{\rm a}}]_{{\rm flux},i} + \sigma^{2} [{\cal I_{\rm a}}]_{{\rm wavelength},i},
\end{equation}
where ${\cal I_{\rm a}}$ refers to any of the indices.

However, unknown sources of random errors may be present in the
data. Following the method described in Gonz\'alez (1993), we checked
whether the standard deviation of index measurements of stars with
multiple observations within the same run was significantly larger
than the expected error, $\sigma$$[{\cal I_{\rm a}}]_{{\rm
expected}}$, for the whole run (using a $F$-test of variances with a
significance level $\alpha$ = 0.1). When this occurs, a residual
random error $\sigma$$[{\cal I_{\rm a}}]_{{\rm residual(1)}}$ was
derived and introduced in the final random error of each star,
$\sigma$$[{\cal I_{\rm a}}]_{{\rm random},i}$, through a
multiplicative factor $f$
\begin{equation} 
\label{factorf}
\sigma [{\cal I_{\rm a}}]_{{\rm random},i} = f~\sigma [{\cal I_{\rm a}}]_{{\rm expected},i},
\end{equation}
which can be computed as
\begin{equation} 
f \simeq \sqrt{1 + \frac{\sigma^{2}[{\cal I_{\rm a}}]_{{\rm
residual(1)}}}{\sigma^{2}[{\cal I_{\rm a}}]_{{\rm expected}}}},
\end{equation}
(see Appendix~B for a detailed justification of the above
equations). The multiplicative factor for each run was computed by
using the index CaT. It was only statistically significant for some
runs, being always $f<1.7$. In particular, run 2 showed an exceptional
agreement between the expected and measured CaT errors and no
correction was applied.

In the following, a detailed description of the mentioned sources of
random errors is given:

{\it (i) Photon statistics and read-out noise.} With the aim of
tracing the propagation of photon statistics and read-out noise, we
followed a parallel reduction of data and error frames with the
reduction package \reduceme\,, which creates error frames at the
beginning of the reduction procedure and translates into them, by
following the law of combination of errors, all the manipulations
performed over the data frames. In this way, the most problematic
reduction steps (flat-field and distortion corrections, wavelength
calibration, sky subtraction, etc) are taken into account and,
finally, each data spectrum has its corresponding error spectrum which
can be used to derive reliable photon errors in the index,
$\sigma$$[{\cal I_{\rm a}}]_{{\rm photon}}$. Typical errors for the
whole sample of stars due to photon noise are $\sigma$$[{\rm
CaT}^{*}]_{{\rm photon}}$ = 0.158, $\sigma$$[{\rm CaT}]_{{\rm
photon}}$ = 0.222 and $\sigma$$[{\rm PaT}]_{{\rm photon}}$ = 0.140. A
detailed description of the estimation of random errors in the
measurement of classical line-strength indices is shown in CGCG, 
whereas an analogous treatment for the new generic indices
defined in this paper is given in Appendix~A2. More details concerning
the sensitivity of different indices to the signal-to-noise ratio have
already been explained in Section~\ref{sensitivity}.

{\it (ii) Flux calibration.} During each observing night, we observed a sample
of spectrophotometric standard stars (from Oke 1990) and derived a flux
calibration curve for each one of them. Within the same run, the indices were
measured using a final flux calibration curve which was computed by averaging
all the individual ones. In addition, we used the individual curves to
estimate the random error in the flux calibration as the r.m.s scatter among
the different index values obtained with each one of them. Such an error
depends on the observing run, but typical values introduced by this
effect are $\sigma$$[{\rm CaT}^{*}]_{{\rm flux}}$ = 0.033, $\sigma$$[{\rm
CaT}]_{{\rm flux}}$ = 0.059 and $\sigma$$[{\rm PaT}]_{{\rm flux}}$ = 0.057,
being almost negligible in comparison with the photon errors.

\begin{figure*}
\centerline{\hbox{
\psfig{figure=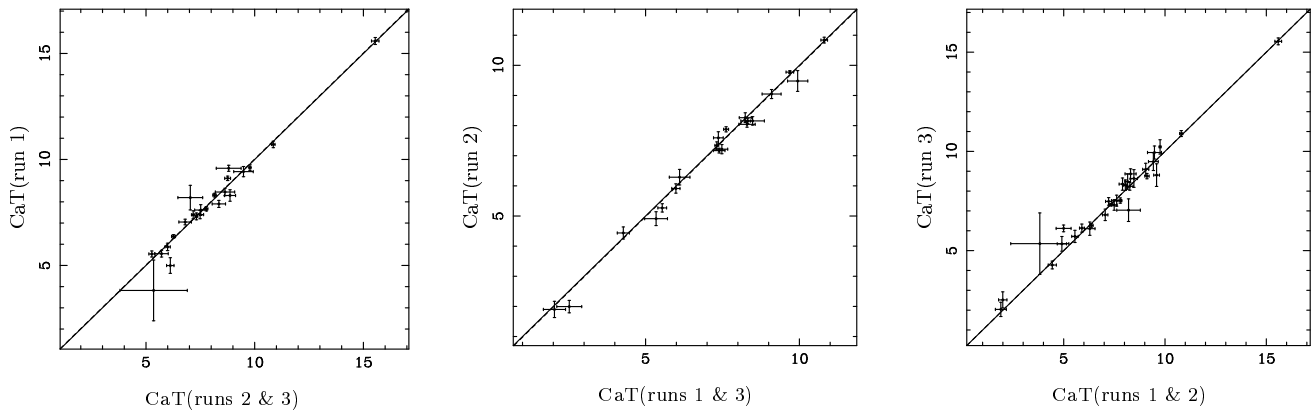}
}}                                                   
\caption{Comparison of CaT measurements of stars with observations in
the JKT runs (1, 2 and~3). The solid line shows the error-weighted
derived offset, whereas the dashed line (almost hidden by the former
fit) displays the one-to-one relation. Error bars for each point are
also plotted.}
\label{comp123}
\end{figure*}

{\it (iii) Wavelength calibration and radial velocity.} The combined
effect of wavelength calibration and radial velocity correction is
another random error source. Radial velocities for Coma, Hyades and
field stars were taken from the Bibliography of Stellar Radial
Velocities (Abt \& Biggs 1972) and from the Hipparcos Input Catalogue
(Turon et al. 1992), the later giving mean probable errors of $\sim
5$~km~s$^{-1}$. The adopted values for the cluster stars were the
averaged cluster radial velocities: M3, M5, M10, M13, M71, M92 and
NGC6171 were taken from Hesser, Shawl \& Meyer (1986), NGC188 from
Friel (1989), and M67 and NGC7789 from Friel \& Janes (1993). Typical
radial velocity errors for the cluster stars are $\simeq
15$~km~s$^{-1}$ ($\sim$ 0.43~\AA\ at $\lambda$~8600~\AA). Once the
wavelength calibration procedure was applied to the whole star sample,
we still found small shifts (a typical value is 0.32 pixels -- or
9.5~km~s$^{-1}$ -- for the JKT runs) between spectra within the same
run, and also systematic offsets (an averaged value is 0.51 pixels)
between different runs. The deviations within each run were probably
due to uncertainties in the published radial velocities and random
locations of the star across the spectrograph slit, along the spectral
direction. In order to correct for these effects, we cross-correlated
each library star with an accurately calibrated reference spectrum (a
hot or cool library star depending on the spectral type of the problem
spectrum). The derived shifts were applied to the corresponding
spectra obtaining thus a highly homogeneous wavelength scale for the
whole stellar library. In spite of that, the indices were measured
assuming a radial velocity error of 5~km~s$^{-1}$ which translates
into rather small errors, typically $\sigma$$[{\rm CaT}^{*}]_{{\rm
wavelength}}$ = 0.003, $\sigma$$[{\rm CaT}]_{{\rm wavelength}}$ =
0.015 and $\sigma$$[{\rm PaT}]_{{\rm wavelength}}$ = 0.003.

To summarize, photon noise is the main source of random error in the
measurement of the present calcium indices. Note, however, that
systematic effects such as an unproper flux calibraton (see
Section~\ref{fluxcalx}) can be the dominant source of uncertainty.

\subsection{Systematic effects}
\label{systematic}

The main sources of systematic effects in the measurement of spectral
indices are spectral resolution, sky subtraction and flux calibration.

{\it (i) Spectral resolution.} As it is shown in
Section~\ref{sensitivity}, the new indices are quite insensitive to
broadening within an ample range of FWHM values. However, in order to
obtain an homogeneous spectral library for stellar population
synthesis models, we broadened the spectra in runs 1, 3 and 4 to the
spectral resolution of run 2 (1.50~\AA). Only the cluster stars
corresponding to runs 5 and 6 (12 and 15 stars, respectively) were not
broadened. These stars keep the spectral resolution given in
Table~\ref{instrum}.

{\it (ii) Sky subtraction.}  The exposure times of most of the library
stars are short enough to neglect the effect of an uncertain
determination of the sky level. However, some faint cluster stars
needed up to 1800 seconds of exposing time, and many of those in
globular clusters were observed with the unavoidable presence of
cluster members inside the spectrograph slit which complicated the
determination of the sky regions. Although there is not a simple
recipe to compute the relevance of this systematic effect, we are
confident that the sky subtraction is roughly correct since we do not
detect unremoved sky lines. In any case, we have implicity quantified
the sky subtraction uncertainties, as well as that of all the
remaining untreated systematic effects, by comparing the amplitude of
the final random errors of the same stars in different runs. This
procedure is explained in greater detail below (Section~\ref{sec33}).

{\it (iii) Flux calibration.} Systematic errors are expected to be
present due to possible differences between the spectrophotometric
system of each run. In order to guarantee that all the spectra are in
an homogeneous system, we selected run 1 as our spectrophotometric
reference system (since it contains a large number of stars in common
with other runs) and derived a re-calibration curve for each other run
by averaging the ones obtained from stars in common with run 1. Even
so, it is not completely sure that our data are in the true
spectrophotometric system, so we encourage readers interested in using
our results to include in their observing plan stars in common with
our stellar sample in order to ensure a proper correction of the data.

\subsection{Final errors}
\label{sec33}

In order to check that the systematic effects between runs have been
fully corrected, we have compared the Ca indices for the stars in
common in the three runs at the JKT applying a $t$ test (see
Fig.~\ref{comp123}).  It is apparent that the agreement between runs
is satisfactory.

A new test to improve the random errors derived in
Section~\ref{random} is to compare the index measurements of stars in
common between different runs. Considering again run~1 as our
reference system, the rest of runs were compared with it to prove the
reliability of the previously derived random errors. If the
r.m.s. from the index comparison agreed with the expected error
obtained from their $\sigma$$[{\rm CaT}]_{{\rm random},i}$ values
(using again an $F$-test of variances), the run being considered was
added to the reference system enlarging the fully calibrated set for
the following comparison. If this was not the case, we proceeded as in
Section~\ref{random} and a second residual error $\sigma$$[{\rm
CaT}]_{{\rm residual(2)}}$ was derived and introduced to the total
random error of each star
\begin{equation} 
\sigma[{\rm CaT}]_{i} = g~\sigma[{\rm CaT}]_{{\rm random},i},
\end{equation}
where $g$ is now computed as
\begin{equation} 
g \simeq \sqrt{1 + \frac{\sigma^{2}[{\rm CaT}]_{{\rm
residual(2)}}}{\sigma^{2}[{\rm CaT}]_{{\rm expected}}}}.
\end{equation}
Each time the problem run required a $g$ factor, the whole comparison
procedure was repeated. Fortunately, that was only the case of run 6
($g$ = 1.65) which converged in one iteration, and no additional
corrections were needed for the other runs.

\section{Final indices and the spectral database}
\label{database}

After converting the spectra to the same resolution and wavelength
scale, final spectra for the repeated stars were obtained through an
error-weighted sum of the individual spectra. An electronic database
containing the final spectra of the 706 stars of the stellar library
is available at the URL addresses:\\{\tt
http://www.ucm.es/info/Astrof/ellipt/CATRIPLET.html}\\ and\\{\tt
http://www.nottingham.ac.uk/\~{}ppzrfp/CATRIPLET.html}

To summarize, the spectral library spans the range from 8348.85 to
9019.50~\AA\, with a spectral resolution of 1.50~\AA\ (FWHM) although
a few cluster stars from runs 5 and 6 keep the original FWHM given in
Table~\ref{instrum}. The above web page also provides an electronic
table listing full information for each star. It includes the indices
CaT$^{*}$, CaT and PaT measured over the final spectra as well as
their corresponding errors. The Henry Draper Catalogue number, other
names (mainly HR and BD numbers), coordinates (R.A. and Dec), spectral
type, luminosity class, apparent magnitude and atmospheric parameters
(that will be derived in Paper~II) are also given. This electronic
table is also available at the CDS via\\ {\tt
http://cdsweb.u-strasbg.fr/Cats.html}

\section{Summary}

We present a new stellar library in the near-IR spectral region
covering the range $\lambda\lambda$~8348-9020~\AA\ at a spectral
resolution of 1.5~\AA. It consists of 706 stars with atmospheric
parameters in the range 2750~K $< {\it T}_{{\rm eff}} <$ 38400~K, 0.00
$< \log g <$ 5.12~dex, and $-3.45 < {\rm [Fe/H]} < +0.60$~dex (see
Paper II). The aim of this library is to obtain an accurate empirical
calibration of the behaviour of the Ca\,{\sc ii} triplet in individual
stars (see the derived fitting functions in Paper III), as well as to
derive reliable predictions of the Ca strength and spectral energy
distribution of stellar populations in a wide range of ages and
metallicities (see Paper IV).  For this purpose, we have defined a new
set of line-strength indices, namely CaT$^{*}$, CaT and PaT, which
overcome some of the limitations of previous Ca index definitions in
this spectral range. In particular, the CaT$^{*}$ index is specially
suited to remove the contamination from Paschen lines in the
integrated spectra of galaxies. Also, the different sources of random
and systematic index errors in these and previous indices have been
analyzed in detail.

In Section~\ref{calibrations} we give some recipes for those readers
interested in using our system or in converting their own calcium
index data into the new CaT indices.  Final spectra for the 706 stars
of the stellar library and an electronic table listing the index
measurements, errors and any other information for each star are
available at the web pages given in Section~\ref{database}. At the
same location we also include a set of subroutines to compute the new
indices, together with their random errors.

\section*{ACKNOWLEDGMENTS}

We thank Guy Worthey for providing us with a list of candidate stars
at the beginning of this project. Ayvur Peletier's help is
acknowledged with the observations and the web page design. AC
acknowledges the hospitality of the Department of Physics (University
of Durham) and the School of Physics and Astronomy (University of
Nottingham). We also thanks S. Pedraz and A. Gil de Paz for helpful
suggestions.  The JKT, INT and WHT are operated on the island of La
Palma by the Royal Greenwich Observatory at the Observatorio del Roque
de los Muchachos of the Instituto de Astrof\'{\i}sica de Canarias. The
Calar Alto Observatory is operated jointly by the Max-Planck Institute
f\"{u}r Astronomie, Heidelberg, and the Spanish Comisi\'{o}n Nacional
de Astronom\'{\i}a.  This research has made use of the Simbad database
(operated at CDS, Strasbourg, France), the NASA's Astrophysics Data
System Article Service, and the Hipparcos Input Catalog. AC
acknowledges the Comunidad de Madrid for a Formaci\'on de Personal
Investigador fellowship. AV acknowledges the support of the PPARC
rolling grant 'Extragalactic Astronomy and Cosmology in Durham
1998-2002' and of a British Council grant within the British/Spanish
Joint Research Programme (Acciones Integradas). This work was
supported by the Spanish Programa Sectorial de Promoci\'on del
Conocimiento under grant No. PB96-610.


\newpage

\appendix

\section{Computation of generic indices and associated random errors}

In this appendix we give full details concerning the definition and
practical measure of a new type of line-strength index, which will be
referred as {\it generic\/} index. In particular, the new indices
defined and introduced in this paper, namely the CaT, PaT and CaT$^*$,
are examples of generic indices.

First, we briefly summarize the main properties of classical
indices. Their generalization into generic indices appears, in a
natural way, when introducing three new requirements in the measure of
spectral features. The necessity of such a change is readily justified
when considering the advantages of the new definition. Finally, we
face the derivation of approximate formulae for the estimation of
random errors from the mean signal-to-noise ratio per \AA.

\subsection{Classical indices}

The aim of the use of line-strength indices is to obtain a quantitative measure
of a spectral signature.  Up to date, most authors have employed line-strength
indices whose definitions are close to the expression of equivalent width,
\begin{equation}
W_{\lambda} ({\rm\AA}) = \int_{\rm line} \left[ 1 - S(\lambda)/C(\lambda)
  \right] \; {\rm d}\lambda,
\label{equivalent_width}
\end{equation}
where $S(\lambda)$ corresponds to the observed spectrum, and $C(\lambda)$ is
the continuum level at the line location obtained by the interpolation of
$S(\lambda)$ in the local continuum. Note that, as it was pointed out by
Geisler (1984) (see also Rich 1988), at low spectral resolution a
pseudo-continuum is measured instead of a true continuum. In order to prevent
subjective determinations of the continuum and absorption regions, {\it
atomic\/} indices have been established by characterizing each index with the
help of three wavelength bandpasses: the relevant spectral feature is covered
by the central bandpass, whereas the other two bandpasses, located at both
sides of the central region, are employed to define the continuum reference
level through a linear interpolation. In particular atomic indices are computed
as (using the notation employed by Gonz\'{a}lez 1993)
\begin{equation}
I_{\rm a}({\rm\AA}) \equiv \int_{\lambda_{c_1}}^{\lambda_{c_2}} 
                 \left[ 1-S(\lambda)/C(\lambda) \right] \; {\rm d}\lambda,
\label{indice_atomico}
\end{equation}
where $\lambda_{c_1}$ and $\lambda_{c_2}$ are the limits of the central
bandpass, $S(\lambda)$ is the observed spectrum, and $C(\lambda)$ is the local
pseudo-continuum, which is derived by
\begin{equation}
C(\lambda) \equiv S_b \frac{\lambda_r - \lambda}{\lambda_r - \lambda_b}
 + S_r \frac{\lambda - \lambda_b}{\lambda_r - \lambda_b},
   \;\;\;\; {\rm where} \\
\end{equation}
\begin{equation}
S_b \equiv  
  \frac{\displaystyle\int_{\lambda_{b_1}}^{\lambda_{b_2}} S(\lambda) \: {\rm d}\lambda}
  {(\lambda_{b_2}-\lambda_{b_1})}, \qquad 
S_r \equiv  
  \frac{\displaystyle\int_{\lambda_{r_1}}^{\lambda_{r_2}} S(\lambda) \: {\rm d}\lambda}
  {(\lambda_{r_2}-\lambda_{r_1})}, \\
\end{equation}
\begin{equation}
\lambda_b  \equiv (\lambda_{b_1}+\lambda_{b_2})/2 , \qquad
\lambda_r  \equiv (\lambda_{r_1}+\lambda_{r_2})/2,
\end{equation}
being $\lambda_{b_1}$, $\lambda_{b_2}$, $\lambda_{r_1}$, and
$\lambda_{r_2}$ the limits of the blue and red bandpasses
respectively. Following the same philosophy, broader spectral lines
and molecular-band features are usually measured in magnitudes using
the so-called {\it molecular\/} indices, which are computed as
\begin{equation}
I_{\rm m}({\rm mag})  \equiv 
  -2.5 \; \log_{10} \left[
     1 - \frac{I_{\rm a}}{\lambda_{c_2}-\lambda_{c_1}}
                    \right].
\label{indice_molecular}
\end{equation}

The error estimate in atomic and molecular indices was extensively
analysed in CGCG (see also Cardiel 1999). These authors derived a set
of accurate formulae for the computation of random errors in
line-strength indices. The applicability of the error formulae depends
on the availability of reliable error spectra. For this purpose, error
frames can be created at the beginning of the reduction procedure and
processed in parallel with data frames. In addition, it is possible to
use simple expressions to estimate the absolute index errors as a
function of the mean signal-to-noise ratio per \AA.  In particular,
the error of the atomic indices can be derived by
\begin{equation}
  \sigma[I_{\rm a}] \approx
  \frac{c_1 - c_2 I_{\rm a}}{\rm{\it SN}(\AA)},
\label{error_atomico_sn}
\end{equation}
whereas for the molecular indices the expression is
\begin{equation}
  \sigma[I_{\rm m}] \approx
  \frac{2.5\;\log_{10}{\rm e}\;\;c_2}{\rm{\it SN}(\AA)} \approx
  \frac{1.086\;c_2}{\rm{\it SN}(\AA)} \approx
  \frac{c_2}{\rm{\it SN}(\AA)},
\label{error_molecular_sn}
\end{equation}
being ${\rm{\it SN}(\AA)}$ the averaged signal-to-noise ratio measured in the
pixels included in the three bandpasses, and $c_1$ and $c_2$ two constants
which depend on the index definition.

\subsection{Generic indices}

The classical index definition can be generalizated by including
several spectral features in a single measurement. In particular it is
possible to define a generic index which can be characterized by an
arbitrary number of continuum and spectral-feature bandpasses. Apart
from this modification, we have also considered two additional
requirements: the possibility of modifying the contribution of each
spectral-feature bandpass (by defining a multiplicative factor for
each spectral feature), and the derivation of the pseudo-continuum
using an error {\it weighted\/} least-squares fit.

Taking all the mentioned requirements into account, the generic index can be
defined as
\begin{equation}
{\cal I}_{\rm a}({\rm\AA}) \equiv \sum_{k=1}^{N_{\rm f}} 
  \left\{ 
    \xi(k) \int_{\lambda_{c_1}(k)}^{\lambda_{c_2}(k)} 
          \left[ 1-S(\lambda)/C(\lambda) \right] \; {\rm d}\lambda
  \right\},
\label{definicion_indice_generico}
\end{equation}
where the subscript ``a'' indicates that this index is the
generalization of the classical atomic index (i.e.\ it is measured in
\AA), $N_{\rm f}$ is the number of spectral-feature bandpasses,
$\xi(k)$ is the multiplicative factor associated to each of these
bandpasses, and $\lambda_{c_1}(k)$ and ${\lambda_{c_2}(k)}$ are the
limits of the $k^{\rm th}$~spectral-feature bandpass. The
pseudo-continuum is assumed to be obtained through an error weighted
least-squares fit to $N_{\rm c}$ continuum bandpasses.

In practice, the integrals in Eq.~(\ref{definicion_indice_generico}) must be
replaced by summations, i.e.
\begin{equation}
{\cal I}_{\rm a}({\rm\AA}) \simeq \Theta \, \sum_{k=1}^{N_{\rm f}} 
  \left\{ 
    \xi(k) \sum_{i=1}^{N(k)} 
          \left[ 1-S(\lambda_{k,i})/C(\lambda_{k,i}) \right]
  \right\},
\label{indice_generico}
\end{equation}
where $\Theta$ is the linear dispersion (in \AA/pixel, assumed to be
constant along the spectrum), $N(k)$ is the number of pixels covering
the $k^{\rm th}$ spectral-feature bandpass, and $\lambda_{k,i}$ is the
central wavelength of the $i^{\rm th}$ pixel in the $k^{\rm th}$
bandpass. The continuum level is derived as (see e.g.\ Bevington 1969)
\begin{equation}
C(\lambda_{k,i})= \alpha_1+\alpha_2\;\lambda_{k,i}
\label{ecuacion_recta}
\end{equation}
with
\begin{equation}
\alpha_1=\frac{1}{\Delta} \left\{
\Sigma_3\;\Sigma_4-\Sigma_2\;\Sigma_5
\right\},
\end{equation}
\begin{equation}
\alpha_2=\frac{1}{\Delta} \left\{
\Sigma_1\;\Sigma_5-\Sigma_2\;\Sigma_4
\right\},
\end{equation}
\begin{equation}
\Delta = \Sigma_1 \; \Sigma_3 - \Sigma_2 \; \Sigma_2,
\end{equation}
where we have defined the following parameters
\begin{equation}
\Sigma_1 \equiv
\displaystyle\sum_{n=1}^{N_{\rm c}} \sum_{h=1}^{M(n)}
  \frac{1}{\sigma^2[{S(\lambda_{n,h})}]},
\label{parametro_sigma_uno}
\end{equation}
\begin{equation}
\Sigma_2 \equiv
\displaystyle\sum_{n=1}^{N_{\rm c}} \sum_{h=1}^{M(n)}
  \frac{\lambda_{n,h}}{\sigma^2[{S(\lambda_{n,h})}]},
\end{equation}
\begin{equation}
\Sigma_3 \equiv
\displaystyle\sum_{n=1}^{N_{\rm c}} \sum_{h=1}^{M(n)}
  \frac{\lambda^2_{n,h}}{\sigma^2[{S(\lambda_{n,h})}]},
\end{equation}
\begin{equation}
\Sigma_4 \equiv
\displaystyle\sum_{n=1}^{N_{\rm c}} \sum_{h=1}^{M(n)}
  \frac{S(\lambda_{n,h})}{\sigma^2[{S(\lambda_{n,h})}]},
\end{equation}
and
\begin{equation}
\Sigma_5 \equiv
\displaystyle\sum_{n=1}^{N_{\rm c}} \sum_{h=1}^{M(n)}
  \frac{\lambda_{n,h}\;S(\lambda_{n,h})}{\sigma^2[{S(\lambda_{n,h})}]},
\label{parametro_sigma_cinco}
\end{equation}
being $M(n)$ the number of pixels covering the $n^{\rm th}$ continuum
bandpass, and $\sigma^2[{S(\lambda_{n,h})}]$ the variance of the
observed spectrum at the central wavelength of the $h^{\rm th}$ pixel
in the $n^{\rm th}$ continuum bandpass.

It is important to note that since, in general, the bandpasses limits
will not be coincident with the wavelengths corresponding to the edges
of each pixel, fractions of pixels must be considered when performing
the summations at the borders of such bandpasses.

Using the same notation, the generalization of the classical molecular index is
straightforward, namely
\begin{equation}
\begin{array}{@{}l@{}}
{\cal I}_{\rm m}({\rm mag})  \equiv \\
  -2.5 \; \log_{10} \left[
     1 - \displaystyle\frac{{\cal I}_{\rm a}}%
      {\sum_{k=1}^{N_{\rm f}} \xi(k)[\lambda_{c_2}(k)-\lambda_{c_1}(k)]}
                    \right],
\end{array}
\label{indice_generico_magnitudes}
\end{equation}
where the subscript ``m'' denotes that this index is measured in magnitudes.

Next, and in order to derive analytical expressions for the
computation of random errors in the generic indices, we follow a
similar treatment to that described in CGCG for the classical
indices. The expected random error in the measurement of a generic
index of the form given in Eq.~(\ref{indice_generico}) can be
expressed as
\begin{equation}
\frac{\sigma[{\cal I}_{\rm a}]}{\Theta} \simeq
\sigma\left[
\sum_{k=1}^{N_{\rm f}} 
  \left\{ 
    \xi(k) \sum_{i=1}^{N(k)} 
          \left[ S(\lambda_{k,i})/C(\lambda_{k,i}) \right]
  \right\}
\right],
\label{indice_over_theta}
\end{equation}
where the right hand term is a function of $2\times\sum_{k=1}^{N_{\rm
f}} N(k)$ variables,
$(\ldots,S(\lambda_{k,i}),\ldots,C(\lambda_{k,i}),\ldots)$, which are
not all independent. In fact, if one assumes that none of the
\mbox{$N_{\rm f} + N_{\rm c}$} spectral-feature and continuum
bandpasses overlap, and taking into account that $C(\lambda)$ is a
linear fit to $S(\lambda)$ in $N_{\rm c}$ continuum bandpasses, the
covariance terms verify
\begin{equation}
 \begin{array}{@{}r@{\;\;}c@{\;\;}l@{}}
  {\rm cov}(S(\lambda_{k,i}),C(\lambda_{m,j})) & = 0 & 
      \;\forall\; k,m,i,j, \\ \noalign{\smallskip}
  {\rm cov}(C(\lambda_{k,i}),C(\lambda_{m,j})) & \ne 0 & 
      \;\forall\; k,m,i,j, \\ \noalign{\smallskip}
  {\rm cov}(S(\lambda_{k,i}),S(\lambda_{m,j})) & = 0 & 
    \left\{
    \begin{array}{l}
      \forall\; k,m,i,j \\
      {\rm except\;for} \; i=j \; {\rm when} \; 
         k=m, \hspace*{-\textwidth}\\
    \end{array}
    \right. \\
 \end{array}
\end{equation}
\begin{displaymath}
 {\rm with}\; \left\{
 \begin{array}{@{}r@{\;}c@{\;}l@{}}
 k,m & \in & [1,N_{\rm f}], \\
 i,j & \in & [1,N(k)].
 \end{array}
 \right.
\end{displaymath}
With these results in mind, and taking partial derivatives in
Eq.~(\ref{indice_over_theta}), it is easy to obtain
\begin{equation}
\begin{array}{@{}l@{}}
\left( \displaystyle\frac{\sigma({\cal I}_{\rm a})}{\Theta} \right)^2 \simeq
\\ \noalign{\medskip}
\displaystyle
\sum_{l=1}^{N_{\rm f}} \sum_{i=1}^{N(l)} \left\{
 \xi^2(l) \;
   \frac{C^2(\lambda_{l,i})\;\sigma^2[{S(\lambda_{l,i})}]+
         S^2(\lambda_{l,i})\;\sigma^2[{C(\lambda_{l,i})}]}%
    {C^4(\lambda_{l,i})}
\right\}+
\hspace*{-\textwidth} \\ \noalign{\medskip}
\displaystyle
\sum_{l=1}^{N_{\rm f}} \sum_{i=1}^{N(l)}
\sum_{m=1}^{N_{\rm f}} \sum_{j=1 \atop j \ne i\;{\rm when}\;m=l}^{N(m)}
 \Biggl\{
 \xi(l)\;\xi(m) \;
 \frac{S(\lambda_{l,i})\;S(\lambda_{m,j})}%
    {C^2(\lambda_{l,i}) \; C^2(\lambda_{m,j})} \times
\hspace*{-\textwidth} \\ \noalign{\medskip}
\displaystyle 
{\rm cov}(C(\lambda_{l,i}),C(\lambda_{m,j}))
\Biggr\}
\end{array}
\label{errores_indices_genericos}
\end{equation}
where $\sigma^2[{C(\lambda_{k,i})}]$ is the variance of the fitted
continuum, evaluated at the $i^{\rm th}$ pixel of the $k^{\rm th}$
spectral-feature bandpass. Remembering that Eq.~(\ref{ecuacion_recta})
is a function of $3\times\sum_{n=1}^{N_{\rm c}} M(n)$ variables,
($\ldots,\lambda_{l,r},\ldots,
S(\lambda_{l,r}),\ldots,\sigma_{S(\lambda_{l,r})},\ldots$), of which
only ($\ldots,S(\lambda_{l,r}),\ldots$) have error, it is possible to
write
\begin{equation}
\sigma^2[{C(\lambda_{k,i})}] = 
 \displaystyle \sum_{l=1}^{N_{\rm c}} \sum_{r=1}^{M(l)} 
   \left(
     \frac{\partial C(\lambda_{k,i})}{\partial S(\lambda_{l,r})}
   \right)^2 \; \sigma^2[{S(\lambda_{l,r})}],
\end{equation}
with
\begin{equation}
 \begin{array}{@{}l@{\;}c@{\;}l}
  \displaystyle
  \frac{\partial C(\lambda_{k,i})}{\partial S(\lambda_{l,r})} & = &
  \displaystyle\frac{1}{\Delta}\; \left(
  \frac{1}{\sigma^{2}[{S(\lambda_{l,r})}]} \; \Sigma_3 \;
  - \;
  \frac{\lambda_{l,r}}{\sigma^{2}[{S(\lambda_{l,r})}]} \; \Sigma_2 \right) \;
  + \hspace*{-\textwidth} \\ \noalign{\medskip}
  & &
  \displaystyle\frac{\lambda_{k,i}}{\Delta}\; \left(
  \frac{\lambda_{l,r}}{\sigma^{2}[{S(\lambda_{l,r})}]} \; \Sigma_1 \; 
  - \;
  \frac{1}{\sigma^{2}[{S(\lambda_{l,r})}]} \; \Sigma_2 \; \right).
  \hspace*{-\textwidth} \\
 \end{array}
\end{equation}
The non-null covariance terms in Eq.~(\ref{errores_indices_genericos}) can 
be derived using the definition of covariance
\begin{equation}
\begin{array}{@{}l@{}}
{\rm cov}(C(\lambda_{k,i}),C(\lambda_{m,j})) = \\
\noalign{\medskip}
\langle C(\lambda_{k,i}) \; C(\lambda_{m,j}) \rangle -
\langle C(\lambda_{k,i}) \rangle \; \langle C(\lambda_{m,j}) \rangle = \\
\noalign{\medskip}
[
  \langle \alpha_1 \alpha_1 \rangle -
  \langle \alpha_1 \rangle \langle \alpha_1 \rangle
]\;+ 
\\ \noalign{\medskip}
[
  \langle \alpha_1 \alpha_2 \rangle -
  \langle \alpha_1 \rangle \langle \alpha_2 \rangle
] \;
(\lambda_{k,i}+\lambda_{m,j})
\;+ 
\\ \noalign{\medskip}
[
  \langle \alpha_2 \alpha_2 \rangle -
  \langle \alpha_2 \rangle \langle \alpha_2 \rangle
] \;
\lambda_{k,i}\;\lambda_{m,j}.
\end{array}
\label{terminos_covariantes}
\end{equation}
Using the parameters defined in
Eqs.~(\ref{parametro_sigma_uno})--(\ref{parametro_sigma_cinco}), and taking
into account that their variances verify
\begin{equation}
\begin{array}{@{}l@{}}
\sigma^2[\Sigma_1] = \sigma^2[\Sigma_2] = \sigma^2[\Sigma_3] = 0, \\
\noalign{\medskip}
\sigma^2[\Sigma_4] = \Sigma_1,\;\;\hbox{and}\;\; \sigma^2[\Sigma_5] = \Sigma_3,
\end{array}
\end{equation}
is not difficult to show that
\begin{equation}
\langle \alpha_1 \alpha_1 \rangle -
\langle \alpha_1 \rangle \langle \alpha_1 \rangle =
\frac{1}{\Delta^2} [
\Sigma_1 \; \Sigma_3 \; \Sigma_3 - \Sigma_2 \; \Sigma_2 \; \Sigma_3
],
\end{equation}
\begin{equation}
\langle \alpha_1 \alpha_2 \rangle -
\langle \alpha_1 \rangle \langle \alpha_2 \rangle =
\frac{1}{\Delta^2} [
\Sigma_2 \; \Sigma_2 \; \Sigma_2 - \Sigma_1 \; \Sigma_2 \; \Sigma_3
],
\end{equation}
\begin{equation}
\langle \alpha_2 \alpha_2 \rangle -
\langle \alpha_2 \rangle \langle \alpha_2 \rangle =
\frac{1}{\Delta^2} [
\Sigma_1 \; \Sigma_1 \; \Sigma_3 - \Sigma_1 \; \Sigma_2 \; \Sigma_2
].
\end{equation}

On the other hand, the expected random error in generic indices measured in
magnitudes, Eq.~(\ref{indice_generico_magnitudes}), can be computed with
\begin{equation}
\displaystyle\frac{\sigma[{\cal I}_{\rm m}]}{\sigma[{\cal I}_{\rm a}]} = 
\displaystyle
  \frac{2.5\;\log_{10}{\rm e}}{10^{\displaystyle-0.4\;{\cal I}_{\rm m}}
  \; \sum_{k=1}^{N_{\rm f}} \left\{\xi(k) \; 
    [\lambda_{c_2}(k)-\lambda_{c_1}(k)]\right\}}.
\end{equation}

\begin{figure}
\centerline{\hbox{
\psfig{figure=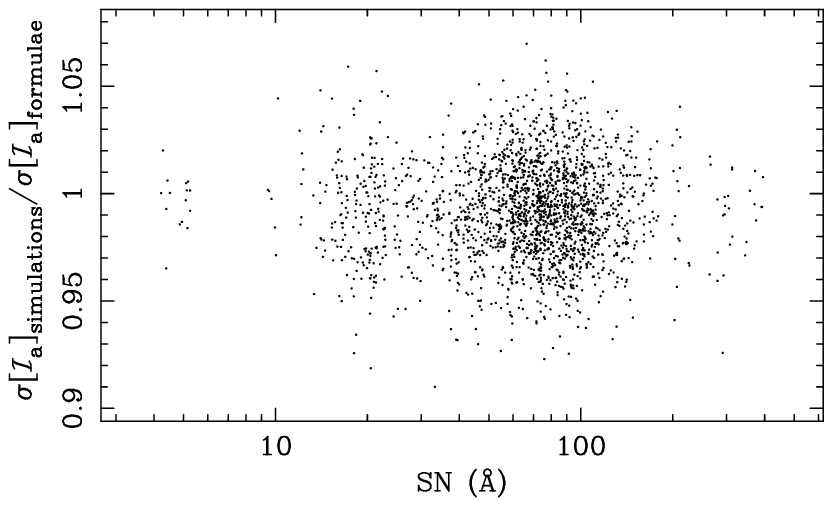}
}} 
\caption{Comparison of absolute random errors in the measure of CaT, PaT and
CaT$^{*}$ in the 706 stars of the library, obtained with the new formulae and
with numerical simulations (1000 simulations per star and index were
performed), as a function of the mean signal-to-noise ratio per \AA\ in the
bandpasses of each index. There is an excellent agreement, within statistical
errors, between both methods (the mean $\sigma[{\cal I}_{\rm a}]_{\rm
simulations}/\sigma[{\cal I}_{\rm a}]_{\rm formulae}$ value for the 2118 points
is 0.99, with a r.m.s.\ of 0.02).}
\label{figura_errores_simulaciones}
\end{figure}

We have verified the reliability of the new set of formulae by comparing the
measured CaT, PaT and CaT$^{*}$ random errors in the 706 stars of the library,
with the errors derived with the help of numerical simulations (see Section~4
in CGCG). This comparison, which is graphically displayed in
Fig.~\ref{figura_errores_simulaciones}, indicates that agreement is complete.
Note, however, that the use of analytical formulae instead of numerical
simulations should be preferred since the later constitutes a computer-time
demanding method.

When the previous formulae are applied over spectra corresponding to redshifted
objects, it is important to remember that generic indices of the form given in
Eq.~(\ref{definicion_indice_generico}), and their errors, must be
corrected to rest frame values by
\begin{equation}
\begin{array}{@{}r@{\;}c@{\;}l}
\left. {\cal I}_{\rm a} \right|_{z=0} & = & \displaystyle
  \frac{1}{1+z} \; \left. {\cal I}_{\rm a} \right|_{z}, \\
\noalign{\medskip}
\sigma[{\cal I}_{\rm a}]_{z=0} & = & \displaystyle
  \frac{1}{1+z} \; \sigma[{\cal I}_{\rm a}]_{z}, \\
\end{array}
\label{index_variation_redshift}
\end{equation}
where $\left. {\cal I}_{\rm a} \right|_{z}$ and $\left. {\cal I}_{\rm a}
\right|_{z=0}$ are the redshifted and rest frame indices, $\sigma[{\cal I}_{\rm
a}]_{z}$ and $\sigma[{\cal I}_{\rm a}]_{z=0}$ are their corresponding errors,
and $z$ is the redshift. Generic indices of the form given by
Eq.~(\ref{indice_generico_magnitudes}), in magnitudes, and their errors, 
do not require correction by redshift.

It is important to note that an appropiate treatment of covariance terms is
essential in order to guarantee the accuracy of the error
estimate. Neglecting such terms can lead to important systematic
anomalies in the determination of random errors. For illustration, we
have studied this effect on the generic indices CaT, PaT and
CaT$^{*}$. For this purpose, we measured the individual absorption
features of CaT and PaT using a new set of indices, CaT$_1$, CaT$_2$
and CaT$_3$, and PaT$_1$, PaT$_2$ and PaT$_3$, respectively, which
were defined as generic indices with the same continuum bandpasses
that the CaT and PaT indices. In this way, the CaT, PaT and CaT$^{*}$
are obviously derived as a linear combination of the new set, i.e.
\begin{equation}
\begin{array}{@{}l@{\;}c@{\;}l}
{\rm CaT}     & = & {\rm CaT}_1 + {\rm CaT}_2 + {\rm CaT}_3, \\
\noalign{\medskip}
{\rm PaT}     & = & {\rm PaT}_1 + {\rm PaT}_2 + {\rm PaT}_3, \\
\noalign{\medskip}
{\rm CaT}^{*} & = & {\rm CaT}_1 + {\rm CaT}_2 + {\rm CaT}_3 - \\
\noalign{\medskip}
              &   & 0.93\;({\rm PaT}_1 + {\rm PaT}_2 + {\rm PaT}_3). \\
\end{array}
\end{equation}
Using Eq.~(\ref{errores_indices_genericos}) for the computation of random
errors in CaT$_i$ and PaT$_i$ (with $i=1,2,3$) in the 706 stars of the library,
and adding errors quadratically neglecting covariance terms in the set CaT$_i$,
PaT$_i$, leads to
\begin{equation}
\begin{array}{@{}r@{\;}c@{\;}l}
\sigma[\Sigma_{i=1}^3{\rm CaT}_i] & = & 
  \left( \sum_{i=1}^3 \sigma^2[{\rm CaT}_i] \right)^{1/2}, \\
\noalign{\medskip}
\sigma[\Sigma_{i=1}^3{\rm PaT}_i] & = & 
  \left( \sum_{i=1}^3 \sigma^2[{\rm PaT}_i] \right)^{1/2}, \\
\noalign{\medskip}
\multicolumn{3}{@{}l}{\sigma[\Sigma_{i=1}^3 {\rm CaT}_i - 
            0.93 \Sigma_{i=1}^3{\rm PaT}_i] =} \\
\noalign{\medskip}
\multicolumn{3}{@{\;\;\;\;}l}{
\left( \sum_{i=1}^3 \sigma^2[{\rm CaT}_i] +
    0.93^2 \sum_{i=1}^3 \sigma^2[{\rm PaT}_i] \right)^{1/2}}.
\end{array}
\end{equation}
The application of these approximate formulae, in comparison with the accurate
expression given in Eq.~(\ref{errores_indices_genericos}), gives
\begin{equation}
\begin{array}{@{}l@{\;}c@{\;}l@{\;}l}
\sigma[\Sigma_{i=1}^3{\rm CaT}_i]/\sigma[{\rm CaT}] & = & 0.7865, & 
  {\rm r.m.s.}=0.0042, \\
\noalign{\medskip}
\sigma[\Sigma_{i=1}^3{\rm PaT}_i]/\sigma[{\rm PaT}] & = & 0.9280, &
  {\rm r.m.s.}=0.0037, \\
\noalign{\medskip}
\multicolumn{4}{@{}l}{\sigma[\Sigma_{i=1}^3{\rm CaT}_i -0.93 
 \Sigma_{i=1}^3{\rm PaT}_i]/\sigma[{\rm CaT}^{*}] = 1.1397,} \\
\noalign{\medskip}
& & & {\rm r.m.s.}=0.0078.
\end{array}
\end{equation}
Summarizing, covariant terms, computed as given by
Eq.~(\ref{terminos_covariantes}), must be retained in order to guarantee that
random errors are free from systematic uncertainties.

\subsection{Estimation of random errors from signal-to-noise ratios}

As we have already mentioned, it is possible to obtain estimates of
the absolute errors of the classical indices as functions of the
signal-to-noise ratio, using expressions as simple as
Eqs.~(\ref{error_atomico_sn}) and~(\ref{error_molecular_sn}). These
formulae constitute a useful tool for observation
planning. Unfortunately, the approximations employed to derive such
expressions (see details in Cardiel 1999) can not be used in the case
of generic indices. For this reason we have decided to use an
empirical approach to this problem.

\begin{figure}
\centerline{\hbox{
\psfig{figure=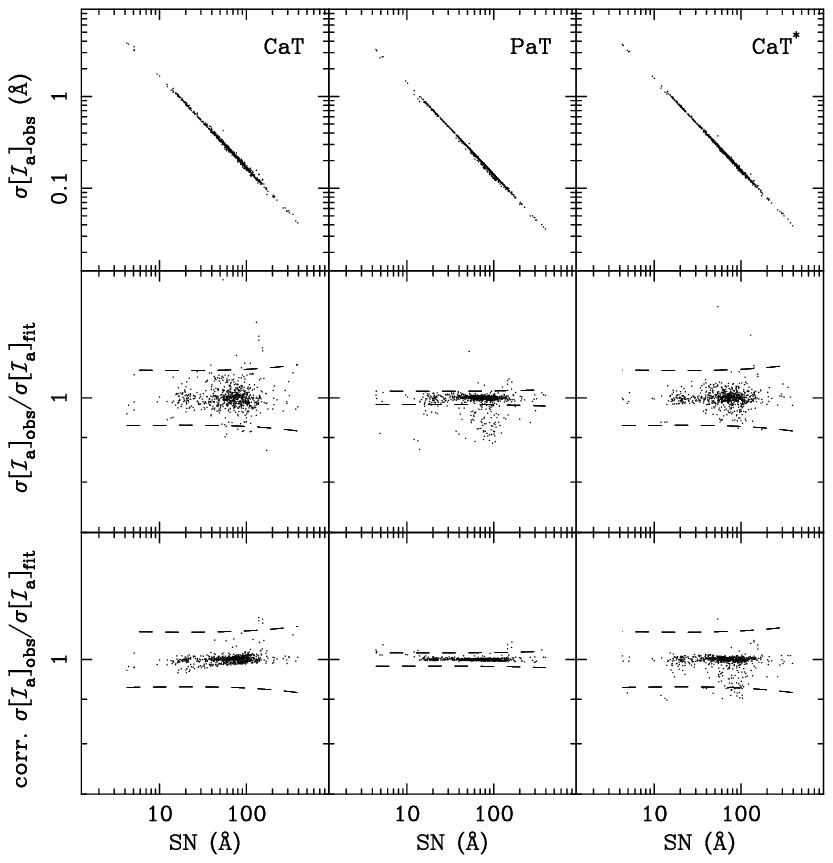}
}} 
\caption{({\it top panels\/}) Absolute errors in the generic indices
CaT, PaT and CaT$^{*}$, measured in the 706 stars of the library, as a
function of the signal-to-noise ratio per \AA. ({\it middle
panels\/}) Residuals of a least-squares linear fit (rejecting data
iteratively outside the 99.73 per cent confidence level ---indicated
by the dashed \mbox{lines---)}.  ({\it bottom panels\/}) Corrected
residuals, obtained by correcting the initial residuals using the
relations given in Eq.~(\ref{relaciones_aproximadas_errores}).}
\label{figura_errores_sn}
\end{figure}

\begin{figure}
\centerline{\hbox{
\psfig{figure=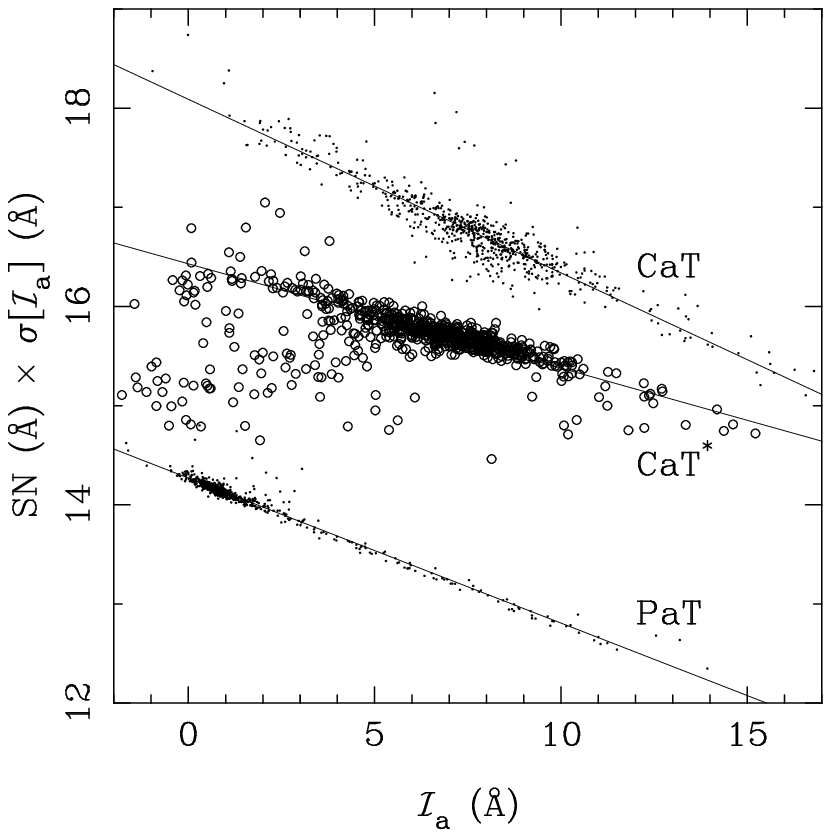}
}} 
\caption{Empirical estimation of the constant factors $c_1$ and $c_2$
of the Eq.~(\ref{relacion_error_sn_buscada}).  The scatter for low
CaT$^{*}$ data are due to stars with comparable CaT and PaT indices,
which translates into relatively small CaT$^{*}$ values, but errors
corresponding to indices with a larger absolute value. The
least-squares fits to straight lines have been performed rejecting
data iteratively outside the 99.73 per cent confidence level.}
\label{figura_errores_derivation}
\end{figure}

In Fig.~\ref{figura_errores_sn} (top panels) we represent the absolute
errors in CaT, PaT and CaT$^{*}$ as a function of the average
signal-to-noise ratio per \AA, for the 706 stars of the library. An
apparently tight correlation is shown by the three indices, as it is
expected. The residuals of a least-squares fit (middle panels) reveal
that the dispersions of errors (r.m.s.) from the linear fit are 3.3,
2.6 and 2.3 per cent for the CaT, PaT and CaT$^{*}$ indices,
respectively. Although this linear behaviour is a good first
estimation of random errors for the generic indices, taking into
account that the error estimation of classical atomic indices depends
on the index value, we have investigated whether a relationship of the
form
\begin{equation}
\sigma[{\cal I}_{\rm a}] \simeq
  \displaystyle \frac{c_1-c_2\;{\cal I}_{\rm a}}{S\!N(\rm\AA)}
\label{relacion_error_sn_buscada}
\end{equation}
can also be employed with the generic indices. For this purpose, we
have represented in Fig.~\ref{figura_errores_derivation} the product
$S\!N(\rm\AA) \times \sigma[{\cal I}_{\rm a}]$ as a function of the index
value, for the stellar library. It is clear that a good linear relation is
exhibited by the three measured indices. The result of a least-squares fit to
the data is
\begin{equation}
\begin{array}{@{}r@{\;}c@{\;}l}
\sigma[\rm CaT(\AA)] & \simeq & 
 \displaystyle\frac{18.09-0.1751\;\rm CaT}{S\!N(\rm\AA)}, \\
\noalign{\medskip}
\sigma[\rm PaT(\AA)] & \simeq & 
 \displaystyle\frac{14.27-0.1463\;\rm PaT}{S\!N(\rm\AA)}, \\
\noalign{\medskip}
\sigma[\rm CaT^{*}(\AA)] & \simeq & 
 \displaystyle\frac{16.43-0.1052\;\rm CaT^{*}}{S\!N(\rm\AA)}. \\
\end{array}
\label{relaciones_aproximadas_errores}
\end{equation}
Since the fitted $c_1$ coefficients are two orders of magnitude larger
than $c_2$, the dependence of generic index errors on the index values
is not strong. However, the application of the relations given in
Eq.~(\ref{relaciones_aproximadas_errores}) reduces the residuals
showed in Fig.~\ref{figura_errores_sn} (bottom panels) for the CaT and
PaT indices (r.m.s.\ 1.3 and 0.4 per cent, respectively), although it
maintains those for the CaT$^{*}$ (r.m.s.\ 2.3 per cent).

The variation of the coefficients $c_1$ and $c_2$ with redshift, for classical
indices, is described by (Cardiel 1999)
\begin{equation}
\begin{array}{@{}r@{\;}c@{\;}l}
\left. c_1 \right|_z & = & (1+z)^{1/2} \left.\;c_1\right|_{z=0}, 
\\ \noalign{\medskip}
\left. c_2 \right|_z & = & (1+z)^{-1/2} \left.\;c_2\right|_{z=0},
\end{array}
\end{equation}
where $\left. c_1 \right|_z$ and $\left. c_2 \right|_z$ are the coefficients to
be employed in spectra with redshift $z$, whereas $\left. c_1 \right|_{z=0}$
and $\left. c_2 \right|_{z=0}$ are the corresponding values at zero redshift.
We have verified that this is also the case for the generic indices CaT, PaT
and CaT$^{*}$ (for this purpose, we translated the 706 stellar spectra of the
library as a function of fictitious redshifts, while maintaining the
signal-to-noise ratio per \AA). Therefore, the inclusion of the redshift effect
in Eq.~(\ref{relacion_error_sn_buscada}) leads to
\begin{equation}
\begin{array}{@{}r@{\;}c@{\;}l}
\sigma[{\cal I}_{\rm a}]_z & \simeq &
  \displaystyle 
  \frac{\left.c_1\right|_z-
    \left.c_2\right|_z\;\left.{\cal I}_{\rm a}\right|_z}{S\!N(\rm\AA)} =
  \\ \noalign{\medskip}
  & & 
  (1+z)^{1/2} \displaystyle
  \frac{\left.c_1\right|_{z=0}-
    \left.c_2\right|_{z=0}\;\left.{\cal I}_{\rm a}\right|_{z=0}}{S\!N(\rm\AA)},
\end{array}
\end{equation}
where we have made use of the index variation with redshift given in
Eq.~(\ref{index_variation_redshift}). Note that although the absolute error
increases with redshift, the contrary is true for the relative error, i.e.
\begin{equation}
\left. \frac{\sigma[{\cal I}_{\rm a}]}{{\cal I}_{\rm a}}\right|_z =
(1+z)^{-1/2}
\left. \frac{\sigma[{\cal I}_{\rm a}]}{{\cal I}_{\rm a}}\right|_{z=0}.
\end{equation}

It is important to highlight that the relations given in
Eq.~(\ref{relaciones_aproximadas_errores})
should only be used to obtain approximate estimates of the generic index
errors, and that they should not be considered as a substitution of the 
accurate formulae described in the previous section. In fact
the application of approximated equations can lead to misleading index
errors (see example in CGCG).

\section{Residual random errors}

When the unbiased standard deviation of index measurements of stars with
multiple observations, $\sigma$$[{\cal I}_{\rm a}]_{{\rm r.m.s.}}$, is
significantly larger than the expected typical error for the whole run
$\sigma$$[{\cal I}_{\rm a}]_{{\rm expected}}$, we derive a residual
random error, $\sigma$$[{\cal I}_{\rm a}]_{{\rm residual}}$ by the
classical quadratic addition of errors
\begin{equation} 
\sigma^{2}[{\cal I}_{\rm a}]_{{\rm residual}} = \sigma^{2}[{\cal I}_{\rm a}]_{{\rm r.m.s.}} - \sigma^{2}[{\cal I}_{\rm a}]_{{\rm expected}}.
\end{equation}
Although a straight procedure would be to add quadratically the residual term
to the random error of each star, we preferred to introduce it through a
multiplicative factor in order to keep the original relative qualities among
the stars, getting at the same time reliable error spectra that can be used
for any index definition. The introduction of such a factor is justified
since, as we have shown in Appendix A3, random errors of the generic indices
CaT, PaT and CaT$^{*}$ can be derived from analytical functions of the index
value and the signal-to-noise ratio (see Eq.~(A37)). Thus, we consider that the
effect of increasing the random errors is equivalent to enlarging the photon
noise $N_{k}$ of each star to a new value $N^{*}_{k}$ by multiplying with a
factor $f$ $>$ 1
\begin{equation} 
N^{*}_{k} = f N_{k}, \;\;\;{\rm and \; therefore}
\end{equation}
\begin{equation} 
{\rm{\it SN}}^{*}_{k}({\rm \AA}) = f^{-1}\; {\rm{\it SN}}_{k}({\rm \AA})
\label{snratio}
\end{equation}
where $k$ refers to the $k^{{\rm th}}$ star. By substituting
Eq.~(\ref{snratio}) in Eq.~(A36), it can be shown that the final
random error, $\sigma$$[{\cal I}_{\rm a}]_{{\rm final},k}$, can be
computed as
\begin{equation} 
\label{sigfsig}
\sigma [{\cal I}_{\rm a}]_{{\rm final},k} \simeq f\;\sigma [{\cal I}_{\rm a}]_{{\rm expected},k}.
\end{equation}
Next, and in order to derive an analytical expression of $f$, the
quadratic addition of errors and Eq.~(\ref{sigfsig}) are simultaneously considered
\[ \sigma^{2}[{\cal I}_{\rm a}]_{{\rm final},k} = \sigma^{2}[{\cal I}_{\rm a}]_{{\rm residual}} + \sigma^{2}[{\cal I}_{\rm a}]_{{\rm expected},k} \simeq \]
\begin{equation} 
\label{finalf}
\simeq f^2\sigma^{2}[{\cal I}_{\rm a}]_{{\rm expected},k},
\end{equation}
which leads to a initial expression of $f$
\begin{equation} 
\label{fexpr}
f \simeq \sqrt{1 + \frac{\sigma^{2}[{\cal I}_{\rm a}]_{{\rm residual}}}{\sigma^{2}[{\cal I}_{\rm a}]_{{\rm expected},k}}}.
\end{equation}
Note that Eq.~(\ref{fexpr}) is not definitive since it depends on the
expected random error of each star. An approximated general expression
is finally obtained by substituting the individual errors by the expected
typical error for the whole run
\begin{equation} 
\label{fexprfinal}
f \simeq \sqrt{1 + \frac{\sigma^{2}[{\cal I}_{\rm a}]_{{\rm residual}}}{\sigma^{2}[{\cal I}_{\rm a}]_{{\rm expected}}}}.
\end{equation}

\label{lastpage}

\end{document}